\documentclass[10pt,prd,aps,twocolumn,nofootinbib,superscriptaddress]{revtex4-2}

\usepackage{graphicx}
\usepackage{booktabs}
\usepackage{dcolumn}
\usepackage{bm}
\usepackage{orcidlink}
\usepackage{multirow}
\usepackage{amsmath}
\usepackage{float}
\usepackage{comment}

\newcommand{\fixme}[1]{{\color{red} #1}}
\newcommand{\ignore}[1]{}

\usepackage{xcolor}
\usepackage{graphicx}
\usepackage{csquotes}
\usepackage{amssymb}
\usepackage{array}
\usepackage{enumitem}
\usepackage{pifont}
\usepackage[range-phrase=\textendash, range-units=single]{siunitx}
\let\sun\odot
\DeclareSIUnit\solarmass{\ensuremath{M_\sun}}
\DeclareSIUnit\year{yr}
\DeclareSIUnit\erg{erg}
\DeclareSIUnit\parsec{pc}
\DeclareSIUnit\c{c}

\graphicspath{{figures/}}
\begin{document}

\preprint{APS/123-QED}

\title{Memory effect from neutron star merger counterparts}

\author{T. Brabant\orcidlink{0009-0003-2919-6035}}
\email{tbrabant@uliege.be}
\affiliation{STAR institute, Université de Liège,
Allée du Six Août, 19C
B-4000 Liège, Belgium}
\affiliation{Institut d'Astrophysique de Paris, Sorbonne Université, 98 bd Arago, 75014, Paris, France}
\author{M. Pillas\orcidlink{0000-0003-3224-2146}}
\affiliation{Institut d'Astrophysique de Paris, Sorbonne Université, 98 bd Arago, 75014, Paris, France}
\author{H. Inchauspé \orcidlink{0000-0002-4664-6451}}
\affiliation{Institute for Theoretical Physics, KU Leuven,
Celestijnenlaan 200D, B-3001 Leuven, Belgium}
\affiliation{Leuven Gravity Institute, KU Leuven,
Celestijnenlaan 200D box 2415, 3001 Leuven, Belgium}
\author{Z. Lin}
\affiliation{Sino-French Institute of Nuclear Engineering and Technology, Sun Yat-Sen University, Zhuhai 519082, China}
\author{F. Foucart}
\affiliation{Department of Physics \& Astronomy, University of New Hampshire, 9 Library Way, Durham NH 03824, USA}
\author{M. Bulla}
\affiliation{Department of Physics and Earth Science, University of Ferrara, via Saragat 1, I-44122 Ferrara, Italy}
\affiliation{INFN, Sezione di Ferrara, via Saragat 1, I-44122 Ferrara, Italy}
\affiliation{INAF, Osservatorio Astronomico d’Abruzzo, via Mentore Maggini snc, 64100 Teramo, Italy}

\date{\today}%

\begin{abstract}

The gravitational-wave (GW) memory effect is a prediction of general relativity, characterized by a permanent change in spacetime geometry. Although it has not yet been observed, next-generation detectors such as the Einstein Telescope (ET) and the Laser Interferometer Space Antenna (LISA) are expected to provide the sensitivity required for its first detection. GW memory comprises two distinct contributions: the nonlinear memory, generated by the self-interaction of GWs, and the linear memory, originating from the anisotropic emission of matter and radiation associated with astrophysical transients. In this work, we quantify both nonlinear and linear memory contributions produced during binary neutron star mergers (BNSs), including those associated with GWs, gamma-ray bursts (GRBs), afterglows, neutrinos, dynamical and disk-wind ejecta, and kilonova emission. We evaluate the detectability of the resulting memory signals with ET and LISA through two applications. First, we apply our analysis to the multi-messenger event GW170817 — GRB 170817A — AT2017gfo. We find that the combined memory signal from this event would have been detectable by ET with a signal-to-noise ratio of 10.7 dominated by the nonlinear memory. We also assess the detectability of the GRB memory across a synthetic population of GRBs, and find that only extreme-energy and jointly favorable configurations could produce a detectable signal in ET. \ignore{In ET, detectability requires short prompt durations ($\sim 0.1$~s). In LISA, it requires either an unusually long prompt phase or an extremely short afterglow duration.} Together, these results demonstrate that while next-generation detectors will probe nonlinear memory, observing linear memory from BNS counterparts will remain elusive except in the most favorable scenarios.
\ignore{Together, these results demonstrate that future observatories will be capable of probing GW memory through multi-messenger observations of compact binary mergers.}

\end{abstract}
\maketitle

\section{Introduction}
\label{sec:intro}
The gravitational-wave (GW) memory effect is a prediction of general relativity that manifests as a permanent change in the spacetime metric~\cite{Christodoulou:1991cr, Thorne.nonlinear, Thorne:1987, Zeldovich:1974gvh, Favata_2010_intro}. It consists of two distinct components. The nonlinear memory is generated by the energy carried by the GWs themselves and is therefore an intrinsic feature of all compact binary coalescence (CBC) systems~\cite{Blanchet:2002av, Favata:2009yd}. In contrast, the linear memory arises from the anisotropic emission of unbound matter or radiation~\cite{Thorne:1987}, such as the asymmetric ejecta produced during the post-merger evolution of binary neutron star (BNS) systems~\cite{Bamber_2026, Lopez:2023aja}. While the linear memory is typically weaker than the nonlinear contribution, it carries complementary information about the energy and dynamics of the ejected material. Detecting the GW memory effect is particularly challenging, but not only because the signal is intrinsically weak. The memory amplitude can actually be quite substantial, reaching roughly $15\%$ of the peak oscillatory waveform and $30\%$ or more for highly spinning configurations~\cite{Inchauspe:LISA}. Another obstacle lies in the response of the detectors. In current ground-based interferometers, the test masses are actively restored to their equilibrium positions after each event, effectively removing the permanent displacement~\cite{Lasky_2016, Hubner:2019sly}. Furthermore, because the memory signal builds up slowly at low frequencies without oscillating, frequency-band limited detectors lack sensitivity in this regime and inherently suppress the memory imprint~\cite{Inchauspe:LISA}.

\ignore{Moreover, the memory amplitude is typically small, reaching up to 10\% of the peak oscillatory waveform, and is concentrated at very low frequencies where detector sensitivity is limited~\cite{Nonlinear_theor, dist_incl_breaking}.} 

Nevertheless, the improved low-frequency performance and enhanced sensitivity of next-generation observatories, such as the Einstein Telescope (ET)~\cite{Hild_2011, Maggiore_2020} and the Laser Interferometer Space Antenna (LISA)~\cite{2017arXiv170200786A,LISA:2024hlh, SciRDv1}, are expected to make the first detection of the GW memory effect feasible. Every GW signal produced by a CBC is expected to be accompanied by a corresponding nonlinear GW memory signal. Since the memory amplitude scales with the total energy radiated during the merger~\cite{Favata_2010_intro}, more massive systems, such as binary black holes, are expected to generate lower frequency memory signals, suiting LISA frequency band well, and have therefore been the primary focus in the literature~\cite{Islo:2019qht, Inchauspe:LISA}. However, BNS mergers offer a unique opportunity to probe both the nonlinear and linear components of the memory effect. In addition to the memory generated by the GW themselves, their rich multimessenger counterparts, including electromagnetic emission~\cite{Metzger:2019zeh, Nakar:2019fza}, expected production of neutrinos~\cite{Metzger:2019zeh, Foucart:2022kon}, and anisotropic matter ejecta~\cite{Metzger:2019zeh}, are predicted to generate an additional linear memory contribution. ET is expected to detect on the order of $10^4$ BNS mergers per year~\cite{Maggiore_2020}, providing an unprecedented sample to study GW memory from BNS and to test the consistency between the observed memory signal and the primary GW emission. Furthermore, although BNS mergers are not expected to be directly observable by LISA through their GW signal, the memory signal, which accumulates at low frequencies, may offer an alternative solution for their detection in the milli-Hertz band. 

The linear memory effect from BNS mergers has been investigated in several studies, each focusing on specific emission channels or phases of the merger. For example, Bamber et al., 2026~\cite{Bamber_2026} combines the memory contributions from the GW signal together with the neutrino, matter ejecta, and electromagnetic emission produced up to $\sim25$ ms post merger. Sago et al., 2004~\cite{Sago_2004} investigates the memory generated by gamma-ray burst (GRB) jets, while Huang et al., 2023~\cite{EInjection_GRBmem} computes the memory produced by a subset of GRB afterglows showing short flares in their light curves. Finally, Lopez et al., 2024~\cite{Lopez:2023aja} studies the contribution of tidal effects and anisotropic dynamical ejected matter to the memory signal. Although these works provide valuable insights into individual sources of GW memory, none presents a unified framework that combines all of the expected BNS signatures over the full post-merger evolution, on longer timescales. 

Consequently, a comprehensive picture of the total memory signal from a multimessenger BNS event remains lacking. In this work, we combine the nonlinear memory effect from BNS mergers with the linear memory contribution from 1) the dynamical and disk wind ejected matter, 2) the neutrinos emitted by the post-merger remnant, 3)
the GRB prompt and afterglow emission, 4) the kilonova (KN) emission. We first consider the unique multimessenger detection GW170817 - GRB 170817A - AT2017gfo~\cite{LIGOScientific:2017ync} as a toy model and explore the detectability of its corresponding memory signal in ET and LISA. Secondly, we focus on a population of GRBs and assess its detectability through their memory effect in the 3$^{\rm  rd}$ generation (3G) of GW detectors. 

The paper is organized as follows: 
Sec.~\ref{sec:counterpart} gives a presentation of the expected BNS counterparts. Sec.~\ref{sec:nonlinear} and~\ref{sec:linear} describes the mathematical framework used to compute the various nonlinear and linear memory contributions, respectively. The methods for estimating the detectability in ET and LISA are presented in Sec.~\ref{sec:snr_and_case_study} and will be applied in two distinct contexts. Sec.~\ref{sec:170817_case} presents the study case of GW170817 -- GRB 170817A -- AT2017gfo, and Sec.~\ref{sec:GRBs} contains a focus on GRB memory effect applied to a synthetic population, both including an estimation of the detectability. Finally we discuss the results and opens on improvements to this work that could be implemented in Sec.~\ref{sec:discussion}.

\section{Neutron star merger counterparts}
Any isolated system that loses energy anisotropically creates a memory effect~\cite{Christodoulou:1991cr, Thorne.nonlinear, Thorne:1987, Zeldovich:1974gvh, Favata_2010_intro}. Such energy can come from the GW energy itself~\cite{Favata_2010_intro} or the energy carried by unbound matter or photons~\cite{Thorne:1987}. A BNS merger is a particularly promising source in this respect~\cite{Bamber_2026}, since it is expected to be accompanied by multiple multimessenger counterparts~\cite{Metzger:2019zeh}. This was confirmed by the historical event of August 17, 2017~\cite{discovery_170817}, which remains the first and, to date, only confirmed detection of a BNS merger with GW and electromagnetic counterparts.

In addition to the GW signal GW170817, detected by the LIGO and Virgo detectors, a short GRB, GRB 170817A, was observed by Fermi/GBM and INTEGRAL/SPI-ACS approximately 1.74~s after the merger~\cite{GRB_detect_170817}. Furthermore, the detection of the kilonova AT2017gfo made it possible to constrain the properties of the associated ejected matter~\cite{KN_estim_170817}. Although theoretically expected, no neutrinos have been detected in association with this event~\cite{neutrino_170817}. In this section, we introduce these counterparts and the relevant quantities to estimate their memory effect.
\label{sec:counterpart}
\subsection{Gravitational waves}
As the BNS system loses energy by emitting GWs, the coalescence proceeds through three phases:
\begin{enumerate} [label=(\Roman*)]
\item Inspiral: the orbit shrinks because the system loses orbital energy.
\item Merger: the two NSs collide and merge at $t=0$, the remnant is either a stable or unstable NS, or it collapses promptly into a black hole.
\item Ringdown: the remnant continues to emit GWs as it settles into a stable configuration.
\end{enumerate}

\begin{figure}[!ht]
    \centering
    \includegraphics[width=1\linewidth]{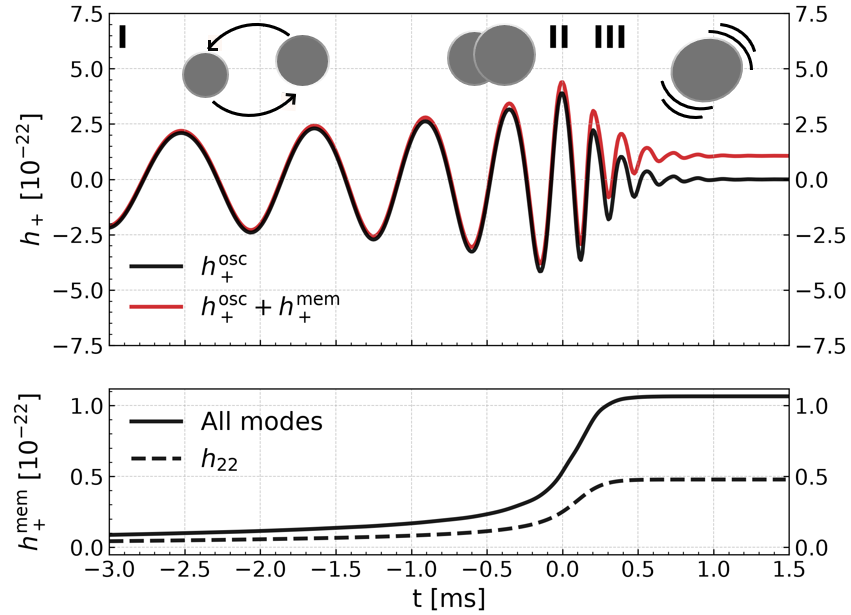}
    \caption{\textit{Top}: oscillatory $h_{+}^{\rm osc}$ (black) and total (red) GW strain, including the memory effect. \textit{Bottom}: isolated memory component $h_{+}^{\rm mem}$. The waveform corresponds to an edge-on equal-mass binary ($M_{\rm tot} = 3~M_\odot$ at $40$~Mpc) obtained with the NRSur7dq2~\cite{NRapprox} approximant for all the available modes (solid) and the ($2,2$) mode (dashed) from the oscillatory waveform.}
    \label{fig:illu_nonlinear}
\end{figure}

These stages are directly imprinted in the GW waveform of the CBC as illustrated in Fig.~\ref{fig:illu_nonlinear}. Throughout these phases, the memory signal (displayed in the bottom panel) builds up gradually, with a slow rise during the inspiral, a steeper slope near the merger and then saturates after the ringdown. The same picture holds for any CBC implying that a nonlinear memory contribution is always present.  
\ignore{The GW signal from a BNS coalescence occurs in three successive phases:
inspiral, merger, and ringdown. Throughout these phase the memory signal builds up, with a steeper slope near the merger as most of the energy is released, and then saturates. More details are given in the following Sec.~\ref{sec:nonlinear} but note that, as the loose of energy by GW is common to all CBCs, the latter constitutes a channel through which a nonlinear memory contribution is always present in such system.}

\subsection{Neutron star ejected matter}
\label{sec:ejecta}

The merger of two NSs expels baryonic matter~\cite{Metzger:2019zeh,1999A&A...341..499R, Bernuzzi_2020, Disk_wind_intro}. The ejected material is commonly classified into two components: the \textit{dynamical ejecta}, from the disruption of the NSs starting during the inspiral phase and the \textit{disk wind ejecta}, a contribution from the post-merger accretion disk. In both cases, the ejected matter is unbound from the system and expelled anisotropically, therefore sources a linear memory effect.

\subsubsection{Dynamical ejecta}
\label{subsub: dyn_intro}
The dynamical ejecta is produced through two distinct processes over a timescale of a few milliseconds~\cite{Nakar:2019fza}. During the late inspiral, the outer layers of NS matter are torn away by tidal forces and expelled preferentially in the equatorial plane~\cite{Radice:2018,Metzger:2019zeh}. Separately, the shock formed at the interface between the two NSs at the moment of contact compresses and squeezes out material, which is ejected preferentially along the polar directions~\cite{Metzger:2019zeh}.

The relative importance of these two mechanisms depends on the properties of the binary. A stiffer nuclear equation of state (EOS) yields less compact NSs that are more easily deformed, therefore enhancing the tidally stripped component~\cite{Radice:2018,Tanaka:2013, Sekiguchi:2016}.  Conversely, a softer EOS yields compact NSs that undergo more violent collision, favoring the shocked polar ejecta. More asymmetric mass ratios similarly tend to eject more mass, since the lighter companion is more strongly tidally disrupted~\cite{Metzger:2019zeh}. If the merger promptly collapses to a black hole, the shock-interface component is instead strongly suppressed, as the material is swallowed by the newly formed remnant~\cite{Metzger:2019zeh, Kawaguchi_2020}. Typical dynamical ejecta masses for BNS mergers range from $10^{-4}$ to $10^{-2}\,M_\odot$, with velocities between $0.1$ and $0.4\,c$~\cite{Nakar:2019fza,Metzger:2019zeh}.

The composition of the dynamical ejecta is characterized by its electron fraction $Y_e$, defined from the local proton and neutron densities, $n_p$ and $n_n$, as~\cite{Metzger:2019zeh}
\begin{equation}
    Y_e = \frac{n_p}{n_n + n_p}.
\end{equation}
This quantity determines whether the ejecta is sufficiently neutron-rich to synthesize heavy nuclei with atomic mass number $A \gtrsim 90$ through the r-process, which proceeds efficiently for $Y_e \lesssim 0.5$~\cite{Metzger:2019zeh, Sekiguchi:2016}. The shock-interface polar component typically has a higher electron fraction and produces elements with $A$ between $90$ and $130$~\cite{Metzger:2019zeh}, resulting in a lanthanide-poor ejecta associated with blue KN emission. By contrast, the tidally stripped component is more neutron-rich ($Y_e \lesssim 0.1$)~\cite{Metzger:2019zeh, Sekiguchi:2016}, undergoes a stronger r-process, and is favorable for producing heavy lanthanide elements~\cite{Sekiguchi:2016}.

\subsubsection{Disk wind ejecta}
\label{subsub: wind_intro}
Part of the neutron-rich matter does not immediately accrete onto the central remnant, but instead circularizes into an accretion disk~\cite{Metzger:2019zeh, Fern_ndez_2013}. This disk launches outflows on a timescale of roughly one second after the merger~\cite{Kasen:2017sxr, Nakar:2019fza}.
Within the first $\sim 100$~ms, intense neutrino emission from the hot remnant drives neutrino-driven winds~\cite{Metzger:2019zeh, Kasen:2017sxr}. Additionally, magnetically driven winds can occur earlier (within the first $0.1$~s) and are generally more dynamically important than neutrino-driven winds~\cite{Kiuchi:2026thl}.  On longer timescales, viscous and magnetic stresses within the disk transport angular momentum outward, causing the disk to spread, heat up, and eventually unbind part of its material as a thermally driven wind, on a timescale of roughly $10$~s~\cite{Metzger:2019zeh, Nakar:2019fza}.

Because it is launched later, the disk wind ejecta trails behind the dynamical ejecta and is distributed more isotropically. Its typical velocity is lower, of order $0.05$ to $0.2\,c$, with masses in the range $10^{-3}$ to $10^{-2}\,M_\odot$~\cite{Nakar:2019fza, Fern_ndez_2013}. Its electron fraction, and therefore its lanthanide content and emission color, depends on the nature of the remnant: a prompt collapse to a black hole yields a low-mass, strongly neutron-rich, red wind~\cite{Metzger:2019zeh}, whereas a longer-lived remnant allows stronger neutrino irradiation to raise $Y_e$ up to $\sim 0.2$~\cite{Fern_ndez_2013}, producing a stronger lanthanide-poor blue wind component~\cite{Metzger:2019zeh, Fern_ndez_2013}.

\subsection{Neutrinos}

Neutrinos play an important role in the evolution of neutron star mergers. They are the main source of cooling of NS remnants and post-merger accretion disks. Absorption and emission of electron (anti)neutrinos also drive changes to the electron fraction of the outflows. This is how the hot dynamical ejecta described in the previous section reaches $Y_e \gtrsim 0.3$, even though the pre-merger neutron stars have $Y_e < 0.1$.

Before merger, neutrino emission is negligible. After merger, we have two potentially important sources of neutrinos: the hot surface of the NS, and the accretion disk surrounding the remnant. The first clearly will only be present for as long as the central object remains a NS. The second requires an accretion rate high enough for neutrino cooling of the accretion disk to be efficient~\cite{De:2020jdt}. For a long-lived NS remnant, the disk emission dominates for $\mathcal{O}(100\,{\rm ms})$, with a total neutrino luminosity of a few times $10^{53} \,{\rm ergs/s}$~\cite{Fujibayashi:2020dvr}. At later time, we expect more symmetrical and dimmer neutrino emission from the NS remnant. For remnants that collapse to a black hole more quickly, bright neutrino emission ($\sim 10^{53-54}\,{\rm erg/s}$) may be observed right before collapse, for a much shorter period of time $\mathcal{O}({\rm ms})$~\cite{Kiuchi:2022nin,Foucart:2022kon,2022EPJA...58...99C}. At later times, the neutrino luminosity drops rapidly~\cite{Kiuchi:2022nin}.

The lifetime of the NS remnant will thus clearly play an important role in setting the neutrino luminosity, and its evolution timescale. For GW170817, that lifetime is uncertain: we know that rapid collapse (on millisecond timescales) is inconsistent with the observed KN signal~\cite{2017ApJ...850L..34B}. A stable remnant may, on the other hand, be difficult to reconcile with the energetics of the ejecta~\cite{2017ApJ...850L..19M}.

\subsection{Gamma-ray burst}

\ignore{Just after the merger of NSs During the merger of NSs, \ignore{bright} GRBs can be emitted. These transient events are among the most energetic phenomena in the Universe~\cite{Fireball_T_Piran, Piran_2005, Kumar:2015}. There are two known progenitors for GRBs: collapses of massive stars~\cite{1993ApJ...405..273W, 1999ApJ...524..262M} and mergers of compact objects involving at least one NS~\cite{1989Natur.340..126E, 2014ARA&A..52...43B, GRB_detect_170817}. The former typically produce long GRB (prompt emission lasting $\gtrsim 2$~s) with softer spectrum, while the latter produce short ($\lesssim 2$~s) and harder GRBs.  \ignore{They are usually classified by the duration of their prompt emission into short ($\lesssim 2$~s) and long ($\gtrsim 2$~s) GRBs,related to NS mergers and collapsars, respectively} Although this classification is often used in the literature~\cite{Kouveliotou:1993}, notable exceptions exist~\cite{Ahumada:2021zrl, Rastinejad_2022}.}

GRBs are among the most energetic transient phenomena in the Universe~\cite{Fireball_T_Piran, Piran_2005, Kumar:2015}. They are known to originate from two main classes of progenitors: a subclass of core-collapse supernovae~\cite{1993ApJ...405..273W, 1999ApJ...524..262M} and CBCs involving at least one NS~\cite{1989Natur.340..126E, 2014ARA&A..52...43B, GRB_detect_170817}. The former typically produce long GRBs (prompt emission lasting $\gtrsim 2$~s) with softer spectra, while the latter generate short ($\lesssim 2$~s) and  harder GRBs~\cite{fermi_duration_distrib}. Although this empirical duration-hardness classification is widely used in the literature~\cite{Kouveliotou:1993}, notable exceptions exist~\cite{gehrels_2006_h4vjh-nxd32,Ahumada:2021zrl, Rastinejad_2022}.

The underlying physics remains incompletely understood but GRBs are believed to be powered by the conversion of the kinetic energy of a highly relativistic outflow, launched by a compact central engine, into radiation~\cite{Piran_2005, Kumar:2015}. Regardless of the progenitor, the fireball model~\cite{Fireball_T_Piran} provides a simple framework for interpreting GRB observations, describing the emission through two successive phases.

\subsubsection{Prompt emission}
\label{subsub: prompt_intro}
The prompt emission is the initial, intense flash of gamma-rays. It likely originates from internal dissipation within the jet such as internal shocks between colliding shells of material or magnetic reconnection events~\cite{Fireball_T_Piran,Piran_2005, Meszaros:1997}. These mechanisms accelerate electrons to relativistic velocities, generating synchrotron radiation~\cite{Piran_2005}. As the expanding fireball becomes optically thin, these gamma-rays escape, producing the observed burst~\cite{Piran_2005}.

The actual energy released during this phase depends on the jet opening angle $\theta_j$, which is poorly constrained. Because we only see the illuminated fraction of the sphere, we characterize the burst by its isotropic-equivalent energy, $E_{\gamma,\rm iso}$ i.e. the energy it would have if it emitted uniformly in all directions~\cite{Piran_2005}. While the assumption of an isotropic energy overestimates the true energy budget by a factor of $ (1-\cos\theta_j)^{-1}$~\cite{Frail_2001}, it provides a convenient, observable scale that we use throughout this work. 

The timescale of this energy release is given by the burst duration $T_{90}$ defined as the time required to accumulate 5\% to 95\% of the total photon counts~\cite{Piran_2005}. Because the prompt phase involves the sudden acceleration and escape of material, both $E_{\gamma,\rm iso}$ and $T_{90}$ govern the amplitude and characteristic frequency of its associated memory signal.

\subsubsection{Afterglow}
\label{subsub: intro_aft}
After the prompt phase, the jet sweeps the surrounding interstellar medium. The collision drives a forward shock that continuously transfers the kinetic energy of the jet to the external gas, decelerating the outflow~\cite{Meszaros:1997}. This process powers the afterglow: a longer-lived, broadband synchrotron emission visible from radio to X-ray~\cite{Meszaros:1997, Sari:break}. Similarly to the prompt emission, we characterize the afterglow using its isotropic-equivalent kinetic energy, $E_{\rm K}$~\cite{Kumar:2015}.

Unlike the prompt flash, the afterglow can last for days, weeks to months~\cite{Meszaros:1997, Fong_2015}, depending on the observing wavelength. To define a consistent, wavelength-independent timescale for the memory signal, we rely on two timescales: the deceleration time $t_{\rm dec}$ and the jet break time $t_{\rm break}$.

The deceleration time $t_{\rm dec}$ marks the onset of the afterglow, when the relativistic ejecta (a fraction $\sim \Gamma^{-1}$ from the ejected material~\cite{Meszaros:1997}) has swept up enough external material to slow down~\cite{Meszaros:1997}, which happens generally on the order of tens to hundreds of seconds~\cite{Kumar:2015, Sari:break}.

The jet break time $t_{\rm break}$ occurs much later, typically after several hours or days~\cite{Sari:break, Rhoads_1999}. Initially, the emission is so highly beamed (Lorentz factor $\Gamma \sim 100 - 1000$~\cite{Fireball_T_Piran, Piran_2005, Ghirlanda_2018, Lithwick_2001}) that the observer only sees a narrow core of the jet, making the outflow appear anisotropic~\cite{Sari:break, Rhoads_1999}. As the blast wave decelerates, its Lorentz factor decreases and the beaming cone widens. Once it becomes wider than the jet actual opening angle, the observer sees past the physical edge of the jet, causing a sharp drop in the observed lightcurve~\cite{Sari:break, Rhoads_1999}. The effect is also independent of the wavelength as it is only a geometric aspect~\cite{Rhoads_1999}. 

Since the emission becomes more isotropic (and weaker) after $t_{\rm break}$, we treat the jet break time as the cutoff for our afterglow memory buildup. Any emission past this time adds negligibly to the overall signal. Thus, $E_{\rm K}$, $t_{\rm dec}$, and $t_{\rm break}$ serve the same modeling purpose for the afterglow that $E_{\gamma,\rm iso}$ and $T_{90}$ serve for the prompt emission. We finally introduce the notation $T_{\rm aft} \equiv t_{\rm break} - t_{\rm dec}$ that refers to the total duration of the afterglow used in this work.

\subsection{Kilonovae}
KNe are optical/infrared transient powered by the radioactive decay of heavy elements synthesized in the neutron-rich ejecta described above~\cite{Metzger:2019zeh}. As the ejected material expands at high velocities, the radioactive heating produces a thermal photon bath that is initially trapped. When the expanding layers cool down and become optically thin, photons finally escape, generating a thermal flux, the KN, that peaks within a few days~\cite{Metzger:2019zeh}. The exact timescale and spectral peak of this emission strongly depend on the ejecta electron fraction: lanthanide-rich material (low $Y_e$) yields a high-opacity, long-lasting red emission, whereas lanthanide-free outflows (high $Y_e$) produce a lower-opacity, faster-peaking blue transient~\cite{Metzger:2019zeh, Jasmin:2026}. Because the distribution of the ejecta is direction-dependent, the escaping photon flux is anisotropic~\cite{Metzger:2019zeh, Jasmin:2026}.

We include KNe as an additional electromagnetic counterpart, as its emission also contributes to the total memory amplitude. Because the photon emission carries away a net momentum flux and since a linear memory effect arises whenever unbound particles are ejected anisotropically, the flux also contributes to the total BNS linear memory. 

\ignore{Several 3D radiative transfer models have been developed to describe KN emission, including for example, \verb|Artis|~\cite{Artis}, and \verb|POSSIS|~\cite{Possis}, the latter being the one employed in this work. POSSIS is a Monte Carlo code that models the time and frequency evolution of photon packets from supernovae or KNe, from 0.1 to 10 days post-merger. Here, we adopt model parameters consistent with those inferred from AT2017gfo. The simulation outputs the energy and frequency $\nu$ of each emitted photon packet for every direction $\theta$ on a skymap at each time step $t$, yielding a specific intensity $F_\nu(t,\theta)$. The time array covers a span starting $0.08$ days after the merger and ending $9.8$ days later, and is uniformly sampled at a frequency of $9.89 \times 10^{-4}\text{ Hz}$. The simulation reveals three distinct phases in the evolution of the KN luminosity, first a bright and strongly polar phase, then a fainter and more isotropic phase, and finally a late phase ($t \gtrsim 5$ days) where the polar and equatorial regions contribute comparably. The corresponding emission spectrum also evolves with time, shifting from short wavelengths ($ \sim 500$ nm) at early times to longer wavelengths ($ \sim 1500$ nm) at later times.}

\section{Nonlinear memory effect}
\label{sec:nonlinear}
In the following section, we provide an overall description of the nonlinear memory. We indicate theoretical approaches to give a clearer picture of the derivation and origin of the nonlinear memory. Then, by manipulating the expression of the memory with spin-weighted spherical harmonics, we show a different formulation that allows for a simple estimate in a non-precessing case and a quasi-circular orbit. Finally, we describe the numerical implementation used in this work and discuss the impact of modes on the amplitude of the memory.   
\subsection{Theoretical framework}
The nonlinear memory effect arises from the quadratic terms in the Einstein field equations, where the energy-momentum carried away by GWs themselves acts as a source for spacetime curvature~\cite{Christodoulou:1991cr, Thorne.nonlinear, Blanchet:2002av, Favata:2009yd, Favata_2010_intro}. Rather than performing the full post-Minkowskian expansion here, we refer the reader to the seminal formulations in~\cite{Christodoulou:1991cr, Thorne.nonlinear, Favata:2009yd}, as well as to Zosso et al. (2026)~\cite{Nonlinear_theor} for a recent derivation based on Isaacson's short-wavelength averaging framework.

In the far-wave zone ($r \gg \lambda_{\mathrm{gw}}$), solving the perturbed field equations with the retarded Green function and projecting onto the transverse-traceless (TT) gauge yields the nonlinear memory strain~\cite{Thorne:1987, Thorne.nonlinear, Christodoulou:1991cr, Favata:2009yd, Favata_2010_intro, GWmemory_text_package}:
\begin{equation}
\delta h^{\mathrm{TT}}_{ij} (t,r) = \frac{4}{r} \int_{-\infty}^{t_{r}} dt' \left[ \int \frac{dE^{\mathrm{gw}}}{dt' d\Omega'} \frac{n_i n_j}{1 - \mathbf{n} \cdot \mathbf{N}} d\Omega' \right]^{\mathrm{TT}},
\label{eq:nonlinear_memory}
\end{equation}
where $t_r = t - r/c$ is the retarded time, $\mathbf{n}$ is the unit vector along the energy emission direction, $\mathbf{N}$ points toward the observer, and $dE^{\mathrm{gw}}/(dt' d\Omega')$ is the GW energy flux per unit solid angle. 

Equation~\ref{eq:nonlinear_memory} also highlights the hereditary nature of nonlinear memory: the metric perturbation accumulates over the past history of the CBC gravitational radiation. As the cumulative energy loss is strictly positive and integrated over the retarded time, it does not oscillate to zero. Instead, it accumulates mostly around merger and leaves a permanent step-like offset in the metric (see Fig.~\ref{fig:illu_nonlinear}).

\subsection{Multipolar decomposition of the memory}
To estimate the nonlinear memory effect, the energy flux can be rewritten in term of the GW polarizations~\cite{Favata:2009yd, Inchauspe:LISA}

\begin{equation}
    \frac{dE^{\mathrm{gw}}}{dt\,d\Omega}
    = \frac{r^2}{16\pi}\left\langle \dot{h}_+^{\,2} +\dot{h}_\times^{\,2}\right\rangle,
    \label{eq: energy_flux_gw}
\end{equation}
\noindent where the brackets mean to average over multiple wavelengths~\cite{Favata:2009yd, Inchauspe:LISA}.

Substituting this expression into Eq.~\ref{eq:nonlinear_memory} leads to a complex integral, because the polarizations depend on both time and angular components. To simplify the double integration,
it is standard in the literature to decompose the complex strain
$h_+ - i h_\times$ into multipoles $h_{\ell m}(t)$ of spin-weighted
spherical harmonics~\cite{Favata:2009yd, Inchauspe:LISA}:
\begin{equation}
    h_+ (t,\theta,\phi) - i h_\times(t,\theta,\phi)
    = \sum_{\ell=2}^{\infty} \sum_{m = -\ell}^{\ell}
      h_{\ell m}(t)\; {}_{-2}Y_{\ell m}(\theta,\phi),
      \label{eq: decomp_h+_hx}
\end{equation}
where ${}_{-2}Y_{\ell m}(\theta,\phi)$ are the spin-weighted spherical harmonics with spin weight $-2$. $\theta$ is the polar inclination angle measured from the positive $z$-axis\footnote{In our work, $\theta = 0$ corresponds to the North Pole (viewing the binary face-on along the positive $z$-axis), $\theta = \pi$ corresponds to the South Pole (face-on along the negative $z$-axis).}, and $\phi$ is the azimuthal angle around the $z$-axis measured from the $x$-axis. For the same reason that there is no GWs emission from a monopole or dipole, the sum over $\ell$ starts at $\ell = 2$. 

Hence, the nonlinear memory can be decomposed in a basis of spin-weighted spherical harmonics. Eq.~\ref{eq:nonlinear_memory} is decomposed in terms of modes ($\ell,m$) by first projecting the energy flux (a scalar quantity) on the basis made of spherical harmonics (of spin 0)~\cite{Favata:2009yd}:
\begin{equation}
    h_{\ell m}^{\rm mem} = \frac{16\pi}{r}\sqrt{\frac{(\ell-2)!}{(\ell +2)!}} \int_{-\infty}^{t_{r}} dt'  \int \frac{dE^{\mathrm{gw}}}{dt' d\Omega'} \bar{Y}_{\ell m}(\Omega),
\end{equation}
\noindent with $\bar{Y}_{\ell m}$, complex conjugate of ${Y}_{\ell m}$.

The energy flux of the GW, composed of the derivatives of the metric (a spin-2 tensorial field), is obtained by projecting Eq.~\ref{eq: energy_flux_gw} on the $_{-2}Y_{\ell m}$~\cite{Favata:2009yd}: 

\begin{equation}
\begin{split}
   \frac{dE^{\mathrm{gw}}}{dt\,d\Omega}= &\frac{r^2}{16\pi} \sum_{\ell'=2}^{\infty} \sum_{\ell''=2}^{\infty} \sum_{m'=-\ell'}^{\ell'} \sum_{m''=-\ell''}^{\ell''}\\
   &\dot{h}_{\ell'm'}\dot{\bar{h}}_{\ell'' m''}
   \, {}_{-2}Y_{\ell' m'}(\Omega) {}_{-2}\bar{Y}_{\ell m}(\Omega).
\end{split}
\end{equation}\\

Hence, the nonlinear memory expression decomposed in spin-weighted spherical harmonics is~\cite{Favata:2009yd,dist_incl_breaking}:

\begin{equation}
\begin{split}
h^{\text{mem}}_{\ell m} = &r \sqrt{\frac{(\ell - 2)!}{(\ell + 2)!}} \sum_{\ell'=2}^{\infty} \sum_{\ell''=2}^{\infty} \sum_{m'=-\ell'}^{\ell'} \sum_{m''=-\ell''}^{\ell''}\\
&\times \mathcal{G}^{\ell \ell' \ell''}_{m m' m"}\int_{-\infty}^{t_r} dt \, \dot{h}_{\ell' m'} \, \dot{\bar{h}}_{\ell'' m''},
\label{eq: nonlinear_mem_compt_modes}
\end{split}
\end{equation}
\noindent with an angular integral $\mathcal{G}_{mm'm''}^{\ell \ell' \ell''}$~\cite{GWmemory_text_package}:
\begin{equation}
\mathcal{G}_{mm'm''}^{\ell \ell' \ell''} = \int d\Omega \, {}_{-2}\bar{Y}^{\ell m}(\Omega) \, {}_{-2}Y^{\ell' m'}(\Omega) \, \bar{Y}^{\ell'' m''}(\Omega).
\label{eq: angular_fac_nonlinear}
\end{equation}\\

Although only a factor $r$ explicitly appears in the Eq.~\ref{eq: nonlinear_mem_compt_modes}, the $r^{-1}$ scaling is preserved  because $h_{\ell m} \propto  r^{-1}$.
\subsection{Simplification for non-precessing quasi-circular binaries}
Under the assumptions of quasi-circular and non-precessing orbits, the nonlinear memory effect can be directly computed from the dominant mode $h_{22}$. Specifically, the memory signal observed with a viewing angle $\theta$, derived from the $h_{22}$ mode only, is given by~\cite{Favata_2010_intro, GWmemory_text_package, Lopez:2023aja}:
\begin{align}
    h_+^{\text{mem}} &= \frac{r}{192\pi} \sin^{2}\theta\,(17 + \cos^{2}\theta) 
        \int_{-\infty}^{t_r} |\dot{h}_{22}|^{2}\,dt, \label{eq:memory_h22+} \\
    h_\times^{\text{mem}} &= 0. \label{eq:memory_h22x}
\end{align}

The polarization axes $x$ and $y$ in the plane of the sky are chosen such that the $x-$axis coincides with the projection of the binary's orbital angular momentum. The cancellation of the $\times$ polarization stems from the assumptions of non-precession and circular orbit in this frame.

\subsection{Numerical computation}
\begin{figure}
    \centering
    \hspace*{-1cm}

    \includegraphics[width=1\columnwidth]{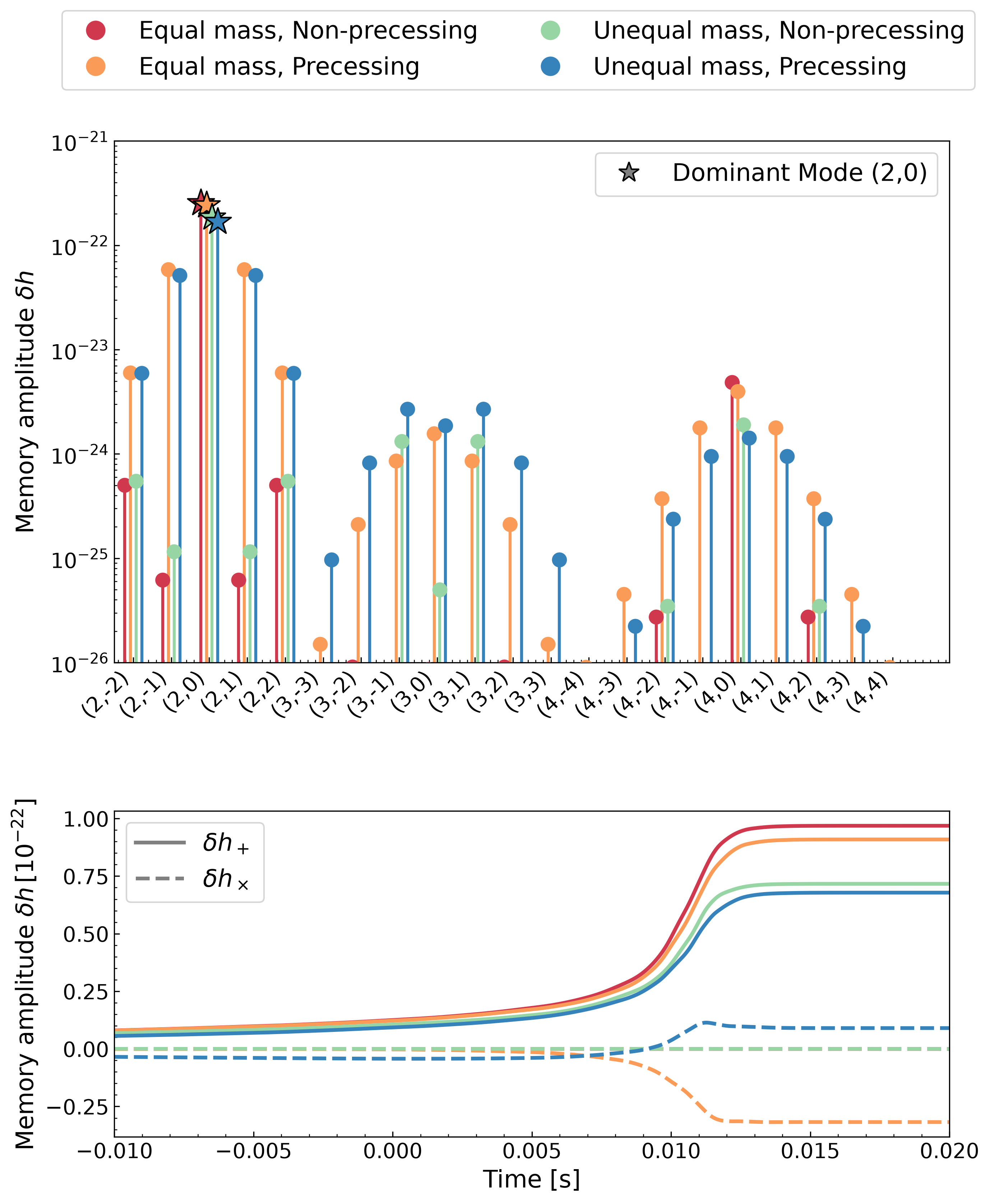}

    \caption{Nonlinear memory effect for a binary system with properties representative of GW170817 ($M_{\rm tot} = 2.73 M_\sun, d = 40~\rm{Mpc}$) modeled with the approximant NRSur7dq2. The plots compare equal ($q =1$) and unequal ($q=2$) mass ratios, both with and without spin-induced precession ($S_{1,x} = 0.5$, $S_{2,y} = 0.5$) seen edge-on. \textit{Top}: Memory amplitude decomposed into spherical harmonics modes ($\ell, m$). \textit{Bottom}: Time evolution of the memory polarizations. Figure inspired by Fig.~1 and Fig.~3 in Talbot et al. (2018)~\cite{GWmemory_text_package}.}
    \label{fig:modes_mem}
\end{figure}

For simplicity, we compute the nonlinear memory effect numerically using the \verb|GWMemory| Python package~\cite{gwmemory_python}, which evaluates Eq.~\ref{eq: nonlinear_mem_compt_modes}. Under the assumptions mentioned above, the memory evolution follows Eq.~\ref{eq:memory_h22+} when only the $h_{22}$ mode is considered. For non-precessing binaries, the memory signal predominantly appears in the $m = 0$ modes (especially $h_{20}$ and $h_{40}$) and is restricted to the $+$ polarization. While unequal-mass systems excite additional $m \neq 0$ modes, the memory is still plus polarized but weaker. The introduction of spin precessing breaks the azimuthal symmetry and leads to a nonzero cross-polarization. The effect on the modes amplitude is represented in Fig.~\ref{fig:modes_mem} for a system similar to GW170817.

Among the models within \verb|GWMemory|, we use the surrogate NRSur7dq2~\cite{NRapprox}. The approximant includes modes up to $\ell = 4$, and provides a rapid estimate of nonlinear memory for systems with a mass ratio less than two~\cite{NRapprox}. A more accurate estimation for BNS should ideally incorporate tidal effects~\cite{Lopez:2023aja}, which are currently absent from models in \verb|GWMemory|. To account for this, one could directly evaluate Eq.~\ref{eq:memory_h22+} using a tidal-inclusive approximant, which could cause slight changes in the results~\cite{Lopez:2023aja} Our estimations based on the analytical integration of the (2,2) mode for a GW170817-like system show that tidal effects estimated with the approximant SEOBNRv4\_ROM\_NRTidalv2~\cite{PhysRevD.100.044003} reduces the amplitude by $\sim 15\%$ near the merger, while comparing the tidal model to NRSur7dq2 yields a $\sim 27\%$ difference. 

\section{Linear memory effect}
\label{sec:linear}
In this section, we review the linear memory effect generated by the multimessenger counterparts of a BNS merger. The designation \textit{linear} stems directly from its derivation. Unlike the nonlinear memory, which requires expanding the Einstein field equations to the second perturbative order~\cite{Christodoulou:1991cr, Thorne.nonlinear}, the linear memory arises from the first-order, linearized field equations. After outlining the theoretical origin of this effect, we detail the two approaches used in this work to model its temporal evolution: the use of analytical toy-models and numerical simulations.
\subsection{Theoretical framework}
The memory effect can also arise under the weak-field assumption of the linearized Einstein field equations, as first shown by Zel'dovich and Polnarev, 1974~\cite{Zeldovich:1974gvh}. They considered a system of $N$ non-interacting point-like particles, each with mass $m_A$ and worldline $x^{\mu}_A(t) = (ct, \mathbf{x}_A)$ where the label $A = 1, ... , N$ distinguishes the individual particles. Their stress-energy tensor is given by Eq.~(2.8.3) in~\cite{Weinberg:1972kfs}:
\begin{equation}
    T^{\mu\nu}(t,\mathbf{x}) = \sum_{A=1}^{N} m_A \gamma_A \frac{dx_A^\mu}{dt} \frac{dx_A^\nu}{dt}\, \delta^{(3)}(\mathbf{x} - \mathbf{x}_A(t)),
    \label{eq: SE_t}
\end{equation}
where $\gamma_A = 1/\sqrt{1 - v_A^2/c^2}$ is the Lorentz factor for a particle of velocity $v_A$ and $dx_A^\mu/dt=  (c,\mathbf{v}_A)$.\\

A permanent effect on spacetime can be demonstrated by substituting Eq.~\ref{eq: SE_t} into the general solution of the linearized Einstein field equations. Using the retarded Green's function for the wave equation, the trace-reversed metric perturbation $\bar{h}_{\mu\nu}$ takes the form
\begin{equation}
    \bar{h}_{\mu\nu}(t,\mathbf{x}) = 4 \int d^3x' \, \frac{T_{\mu\nu}(t_{\mathrm{r}}, \mathbf{x}')}{|\mathbf{x} - \mathbf{x}'|},
\end{equation}
with the stress-energy tensor evaluated at the retarded time $t_r = t - r/c$, where $\mathbf{x'}$ denotes the source coordinates and $r = |\mathbf{x}- \mathbf{x'}|$ is the distance from the source element to the observer located at $\mathbf{x}$.\\

For a distant observer (i.e., $r \gg R_{\rm source}$ where $R_{\rm source}$ is the characteristic size of the source) and working in the transverse-traceless (TT) gauge with natural units, the far-zone metric perturbation for the stress-energy tensor defined in Eq.~\ref{eq: SE_t} reduces to~\cite{Thorne.nonlinear}
\begin{equation}
    h_{ij}^{\mathrm{TT}}(t,\mathbf{x}) = \frac{4}{r} \sum_{A=1}^N \left[ \frac{m_A}{\sqrt{1 - \mathbf{v}_A^2}} \left( \frac{v_A^i v_A^j}{1 - \mathbf{n} \cdot \mathbf{v}_A} \right) \right]^{\mathrm{TT}},
\end{equation}
where $\mathbf{n}$ is the unit vector pointing from the source to the observer.\\

The linear memory effect is then defined as the net change between the final and initial states of the metric~\cite{Favata_2010_intro, Thorne.nonlinear}:
\begin{equation}
    \Delta h^{\mathrm{TT}}_{ij} = \lim_{t \to +\infty} h^{\mathrm{TT}}_{ij}(t,\mathbf{x}) - \lim_{t \to -\infty} h^{\mathrm{TT}}_{ij}(t,\mathbf{x}).
\end{equation}

Thus, the expression of the linear memory effect produced by $N$ particles reads~\cite{Thorne.nonlinear}
\begin{equation}
    \Delta h^{\mathrm{TT}}_{ij}(r) = \frac{4}{r} \, \Delta \sum_{A=1}^N \left[ \frac{m_A}{\sqrt{1 - \mathbf{v}_A^2}} \left( \frac{v_A^i v_A^j}{1 - \mathbf{n} \cdot \mathbf{v}_A} \right) \right]^{\mathrm{TT}},
    \label{eq: final_linear_m}
\end{equation}
where $\Delta$ denotes the difference between the final and initial states of the system.\\

The linear memory can be produced during hyperbolic encounters~\cite{hyperb}, GRBs~\cite{Sago_2004}, or through anisotropic ejection of matter. In such cases, it is referred to as \textit{ordinary} memory (adopting the terminology of Bamber et al., 2026~\cite{Bamber_2026}) as it originates from non-null matter.

\subsection{Modeling the temporal evolution}
\subsubsection{A toy-model approach for the GRBs, dynamical and disk wind ejecta}
\label{subsub: toy_model}
To estimate the temporal evolution of the linear memory effect in a computationally efficient manner, we use a toy model that reproduces the characteristic phases of the memory signal while avoiding the need for costly numerical relativity simulations. The model captures the essential phenomenology of memory buildup through a three-phase evolution: an initial bound phase (I), a rising phase (II), and a final saturated phase (III). The linear memory amplitude $h_+^{\rm mem} (t,r,\theta)$ is then written as

\begin{equation}
h_+^{\rm mem} = \\
    \begin{cases}
        0 & t \le t_{\rm{start}} ~ \rm{(I)} \\[4pt]
        \Delta h(r) \, \Theta(\theta) \, s(t,\tau) & t_{\rm{start}} < t \le \tau + t_{\rm{start}} ~ \rm{(II)} \\[4pt]
        \Delta h(r) \, \Theta(\theta) & t > \tau + t_{\rm{start}} ~ \rm{(III)}
\end{cases}
\end{equation}
where each factor encodes a distinct physical ingredient: $\Delta h(r)$ sets the overall distance-dependent amplitude of the memory (defined in Eq.~\ref{eq: final_linear_m}), $\Theta(\theta)$ describes the angular dependence of the signal, and $s(t,\tau)$ governs how the signal rises in time between the onset of ejection and saturation. The three phases of the model are defined as follows:
\begin{enumerate} [label=(\Roman*)]
    \item $t_{\text{start}}$ marks the onset of mass ejection or acceleration for a specific counterpart. This time is measured relative to the merger time, defined as $t=0$. Before this time, no energy has yet been ejected from the system, so the memory amplitude remains zero. 
    
    \item As unbound material is ejected and accelerated, the memory signal grows in amplitude. The growth is described by the step function $s(t,\tau)$. The parameter $\tau$ corresponds to the duration of the mass-ejection and/or acceleration phase. Physically, this is the timescale over which the bulk of the unbound material achieves its asymptotic velocity, after which no further contribution to the memory amplitude occurs.

    \item Once the ejecta has reached its terminal velocity, the memory amplitude saturates at its maximum value $\Delta h~ \Theta(\theta)$.
\end{enumerate}

The angular dependence of the signal is encoded in the function $\Theta(\theta)$, which accounts for the observer's viewing angle $\theta$ relative to the ejecta symmetry axis. In the coordinate system centered on the merger remnant and aligned with the orbital angular momentum, the angular factor simplifies to~\cite{Mukhopadhyay_2021, Sago_2004}

\begin{equation}
    \Theta(\theta) = \frac{\sin{^2\theta}}{1- \beta \cos{\theta}},
    \label{eq: Theta_def}
\end{equation}
\noindent where $\beta = v/c$, and $\Gamma = (1-\beta^2)^{-1/2}$ is the Lorentz factor of the ejected material. 

The emission exhibits an anti-beaming effect~\cite{Sago_2004}:
\begin{itemize}
    
\item For non-relativistic ejecta ($\beta \ll 1$), $\Theta(\theta)\simeq \sin^2{\theta}$ which is maximized for edge-on viewing ($\theta = \pi/2)$. 

\item For ultra-relativistic ejecta ($\beta \sim 1, \Gamma\gg 1$), the asymptotic form becomes

\begin{equation}
    \Theta(\theta) \simeq \frac{2\theta^2}{\Gamma^{-2} + \theta^2},
\end{equation}
which peaks at $\theta \sim \Gamma^{-1}$~\cite{Sago_2004}: the memory is suppressed along the axis and maximized just outside the relativistic cone of aperture $\sim \Gamma^{-1}$~\cite{Sago_2004}. 
\end{itemize}

The transition of the memory signal from zero to its saturated value can be modeled in several ways, depending on the assumed behavior of the ejecta during this phase. In the following, we consider two models for the step function $s(t,\tau)$:

\begin{equation}
s(t,\tau) =
\begin{cases}
\displaystyle 2\left[\left(1 + e^{-t/\tau}\right)^{-1} - \frac{1}{2}\right] & \text{: exponential} \\[15pt]
\displaystyle \frac{t}{\tau} & \text{: uniform}
\end{cases}
\label{eq:step_function_models}
\end{equation}

The uniform model assumes a constant rate of mass ejection throughout the rise phase, while the exponential model describes a smoother, asymptotic approach to saturation. In the literature~\cite{Sago_2004, EInjection_GRBmem}, the uniform model is predominantly considered over the sigmoid-like evolution~\cite{Yang_2018}. For completeness, we choose to also compare our results against the exponential model.

To investigate the detectability of the signal, its Fourier transform (FT) must be computed. Given the temporal dependence of the memory strain $h(t,r, \theta)$, evaluating the frequency profile of the signal is equivalent to taking the FT of $s(t,\tau)$, defined as $\tilde{s}(f) = \int_{-\infty}^{+\infty} s(t) e^{2\pi i ft}~dt$. Neglecting the Dirac delta distribution at $f=0$, we find the analytical expressions for the FT of each model~\cite{Lopez:2023aja,Sago_2004}:

\begin{equation}
\tilde{s}(f) =
\begin{cases}
\displaystyle \frac{ \tau }{2i \sinh{(\pi f \tau)}} & \text{: exponential} \\[15pt]
\displaystyle \frac{1}{2\sqrt{2 }\pi f^2\tau}[1-\cos{(2\pi f \tau)}]^{1/2} &\text{: uniform }
\end{cases}
\label{eq:FT_step_function_models}
\end{equation}

For both models, we note that $|\tilde{h}(f)| = \Delta h |\tilde{s}(f)| \propto f^{-1}$ in the low-frequency limit ($f \ll 1/\tau$)~\cite{Sago_2004}. This behavior mirrors the FT of a Heaviside step function, which is expected since these low frequencies probe the signal's evolution over timescales much longer than the rising phase $\tau$. To properly assess detectability, it is more convenient to compute the characteristic strain $h_c(f) = 2f|\tilde{h}(f)|$, as it allows for a direct comparison with the sensitivity curve and the signal-to-noise ratio (SNR) in a given detector (see later Sec.~\ref{sec:snr_and_case_study}).

\ignore{A representation of both models in the time and frequency domains is shown in Fig.~\ref{fig:models_phenom_def}.} The frequency behavior of the characteristic strain features two regimes: a flat plateau at low frequencies ($f \lesssim \tau^{-1}$), followed by a power-law decay ($h_c \propto f^{-1}$) at higher frequencies. Hence, the inverse of the characteristic time acts as a knee frequency indicating the maximum frequency at which the memory signal remains optimally detectable (only in term of frequency evolution). However, the two models decay differently beyond the knee frequency: the uniform model exhibits a weaker decay, making it an optimistic scenario for detectability, while the exponential model decays more sharply, making it a more pessimistic scenario for frequencies larger than $\tau^{-1}$.

Both models are used as a phenomenological approach for the linear memory contribution from dynamical, disk wind, and GRB counterparts. 

\subsubsection{Simulation-based temporal evolution for KNe and neutrinos}
\label{subsub: simulations}
In this work, the KN and neutrino memory effects are rather directly investigated with simulations.

We first address the KN simulation. Several 3D Monte Carlo radiative transfer codes have been developed to describe the KN emission, including for example, \verb|Supernu|~\cite{supernu}, \verb|SEDONA|~\cite{sedona},   \verb|ARTIS|~\cite{Artis}, and \verb|POSSIS|~\cite{Possis19,Possis23}, the latter being the one employed in this work. \verb|POSSIS| is a code that propagates Monte Carlo photon packets through expanding ejecta to predict the resulting KN emission. Here, we built a synthetic kilonova using Bulla, 2023 model~\cite{Possis23} that is consistent with AT2017gfo and features two distinct ejecta components: a dynamical ejecta and a disk-wind ejecta component (see Sec.~\ref{sec:ejecta}). We modified the code to output the energy and frequency $\nu$ of each escaping photon packet for every direction $\theta$ on a skymap at each time step $t$, yielding a specific intensity $F_\nu(t,\theta)$. The time array covers a span starting $0.08$ days after the merger and ending $9.8$ days later, and is uniformly sampled at a frequency of $9.89 \times 10^{-4}\text{ Hz}$. The simulation reveals three distinct phases in the evolution of the KN luminosity, first a bright and strongly polar phase, then a fainter and more isotropic phase, and finally a late phase ($t \gtrsim 5$ days) where the emission becomes structured again: polar and equatorial regions contribute comparably, while the flux at intermediate latitudes drops. This drop is shown by the grey patches in Fig.~\ref{fig:comparison_maps} where the local luminosity tends to zero. The corresponding emission spectrum also evolves with time, shifting from short wavelengths ($ \sim 500$ nm) at early times to longer wavelengths ($ \sim 1500$ nm) at later times.

To estimate the memory effect associated with neutrino emission, we base ourselves on detailed distributions of neutrinos taken from recent short simulations of NS mergers using energy-dependent Monte Carlo radiation transport~\cite{Foucart:2026kpd}, complemented by the evolution of the neutrino luminosity observed in a broader range of simulations using gray two-moment transport schemes (e.g.~\cite{Kiuchi:2022nin,2022EPJA...58...99C}) as well as in simulations of post-merger remnants~\cite{Fujibayashi:2020dvr}. The Monte Carlo simulations provide us with the luminosity of neutrinos as a function of time and direction of propagation, thus allowing for a direct calculation of the associated GW memory. Those simulations are however short (up to $10\,{\rm ms}$ post-merger), and limited in their parameter space coverage. The broader range of simulations performed with approximate transport provides us with a better view of neutrino emission over longer time scales and/or a broader range of binary systems.

For the Monte Carlo simulations we first consider a system collapsing $\sim 4\,{\rm ms}$ post-merger (MC-HR simulation from~\cite{Foucart:2024npn}; labeled M140-M130 in~\cite{Foucart:2026kpd}). This is the merger of a $1.4M_\odot$ NS with a $1.3M_\odot$ NS, using the SFHo equation of state. These masses are reasonably consistent with GW170817, but for this mass ratio and equation of state the total ejected mass is $<0.01M_\odot$. Most of the neutrino emission occurs between merger and collapse, and remains dim ($\sim 10^{52}\,{\rm erg/s}$). This is a very pessimistic case for neutrino emission in NS mergers. We also consider the merger of a $1.27M_\odot$ NS with a $1.18M_\odot$ NS, for the same equation of state (i.e. the same mass ratio with both masses reduced by $10\%$). This is simulation M127-M118-Base of~\cite{Foucart:2026kpd}. The simulation ends $10\,{\rm ms}$ after merger, before collapse of the remnant to a black hole. In this case, neutrino emission reaches $\sim 4\times 10^{53}\,{\rm erg/s}$ at the end of the simulation. Neutrino emission is then expected to continue at an elevated level up to collapse or $\mathcal{O}(100\,{\rm ms})$, whichever comes first~\cite{Fujibayashi:2020dvr}. We can thus consider a 'pessimistic' extrapolation of the signal to late time in which neutrino emission suddenly stops at the end of the merger simulation and an 'optimistic' extrapolation in which it continues with the same brightness and angular distribution for $100\,{\rm ms}$. We note that this is optimistic for two reasons: first, the system considered here is likely approaching collapse, and thus brighter than it would be in a more 'steady-state' evolution; second, emission may become more symmetric at later times, reducing the impact of neutrino emission on GW memory.

\subsection{Neutron star ejected matter}

After presenting the general form of the models describing the linear memory effect, we evaluate below the term related to the amplitude of the linear memory effect associated with the two types of ejecta presented in Sec.~\ref{sec:ejecta}.

\subsubsection{Dynamical ejecta}
\ignore{Since the dynamical ejecta is unbound and ejected anisotropically, it produces a linear memory effect.} Following Eq.~\ref{eq: final_linear_m}, the peak memory amplitude can be estimated by modeling the dynamical ejecta as a single, radially ejected point-particle with mass $M_{\rm ej,dyn}$ and velocity $v_{\rm ej,dyn}$, giving~\cite{Lopez:2023aja}
\begin{equation}
\begin{split}
    \Delta h_{\text{dyn}} &= \frac{2 M_{\text{ej,dyn}} v^2_{\text{ej,dyn}}}{r}\\
    & \simeq 3.8 \times 10^{-25} \left(\frac{M_{\text{ej,dyn}}}{0.01 M_\odot}\right)
    \left(\frac{v_{\text{ej,dyn}}}{0.2~ \text{c}}\right)^{2}
    \left(\frac{r}{100~\text{Mpc}}\right)^{-1},
    \label{eq: delta_h_l_dyn}
\end{split}
\end{equation}
where $r$ is the luminosity distance to the source.

It is worth noting that using the full ejected mass $M_{\rm ej,dyn}$ in our point-particle model is a very optimistic upper bound. A more rigorous approach would isolate the asymmetric part of the ejected mass.

The onset of the dynamical ejecta memory is not sharply defined since tidal stripping begins a few milliseconds before merger, and the rise time is short, of order $\tau_{\rm dyn} \sim 1$ ms~\cite{Nakar:2019fza}. We therefore take the start of the signal ($t_{\rm start}$) to be the moment at which the memory reaches half of its value at the time of the merger. This approach does not affect detectability.

\subsubsection{Disk wind ejecta}
\ignore{As with the dynamical ejecta, the disk wind is unbound and therefore also produces a linear memory effect} An estimate of the memory amplitude for the disk wind ejecta, using the same expression as Eq.~\ref{eq: delta_h_l_dyn} with parameters appropriate to the wind component~\cite{Metzger:2019zeh,Nakar:2019fza}, gives
\begin{equation}
\begin{split}
    \Delta h_{\text{wind}} &= \frac{2 M_{\text{ej,wind}} v^2_{\text{ej,wind}}}{r}\\
    & \simeq 9 \times 10^{-26} \left(\frac{M_{\text{ej,wind}}}{0.01 M_\odot}\right)
    \left(\frac{v_{\text{ej,wind}}}{0.1~ \text{c}}\right)^{2}
    \left(\frac{r}{100~\text{Mpc}}\right)^{-1}.
    \label{eq: delta_h_l_wind}
\end{split}
\end{equation}
The disk wind ejecta evolves on longer timescales than the dynamical ejecta, with a characteristic rise time of the order of $\tau_{\rm wind} \sim 1$ to $10$ s~\cite{Nakar:2019fza}.
\ignore{Because the wind evolves more gradually than the dynamical ejecta, its characteristic rise time is longer, of order $\tau_{\rm wind} \sim 1$ to $10$ s.}

\subsection{Neutrinos}

The GW memory triggered by anisotropic neutrino emission was initially studied by \cite{Epstein:1978dv} in the context of core-collapse supernovae, where this effect is comparable in amplitude and energy to the gravitational radiation generated by the fluid motion in the collapse of the supernova core. In BNS mergers, the GW memory is calculated using the same method, which follows the Braginsky–Thorne null-memory form \cite{Thorne:1987}. We use the $h_+$ and $h_\times$ given by \cite{2009ApJ...704..951K}:
\begin{align}
h_{+} &= \frac{2}{r} \int_{0}^{t} dt' \int_{4\pi} d\Omega'\, \frac{dl_{\nu}(\Omega',t')}{d\Omega'} \nonumber \\
&\quad \times (1 + \sin\theta'\cos\phi'\sin\xi + \cos\theta'\cos\xi) \nonumber \\
&\quad \times \frac{(\sin\theta'\cos\phi'\cos\xi - \cos\theta'\sin\xi)^{2} - \sin^{2}\theta'\sin^{2}\phi'}
{(\sin\theta'\cos\phi'\cos\xi - \cos\theta'\sin\xi)^{2} + \sin^{2}\theta'\sin^{2}\phi'}, \label{eq: h_plus_neutrinos} \\
h_{\times} &= \frac{4}{r} \int_{0}^{t} dt' \int_{4\pi} d\Omega'\, \frac{dl_{\nu}(\Omega',t')}{d\Omega'} \nonumber \\
&\quad \times (1 + \sin\theta'\cos\phi'\sin\xi + \cos\theta'\cos\xi) \nonumber \\
&\quad \times \frac{\sin\theta'\sin\phi'\,(\sin\theta'\cos\phi'\cos\xi - \cos\theta'\sin\xi)}
{(\sin\theta'\cos\phi'\cos\xi - \cos\theta'\sin\xi)^{2} + \sin^{2}\theta'\sin^{2}\phi'}. \label{eq: h_cross_neutrinos}
\end{align}

Here primed quantities refer to the source frame, while unprimed quantities are in the observer frame. $\Omega'$ denotes the direction of the neutrino flux, with $\theta'$ the polar angle and $\phi'$ the azimuthal angle; $t'$ is the emission time; $dl_{\nu}/d\Omega'$ is the direction and time dependent neutrino luminosity emitted per unit solid angle; $\xi$ is the viewing angle and $r$ is the luminosity distance. As in \cite{2009ApJ...704..951K}, we choose the $y$ axis in the observer frame to coincide with the $y'$ axis in the source frame, without losing generality.

\begin{figure}
    \centering
    \includegraphics[width=1\linewidth]{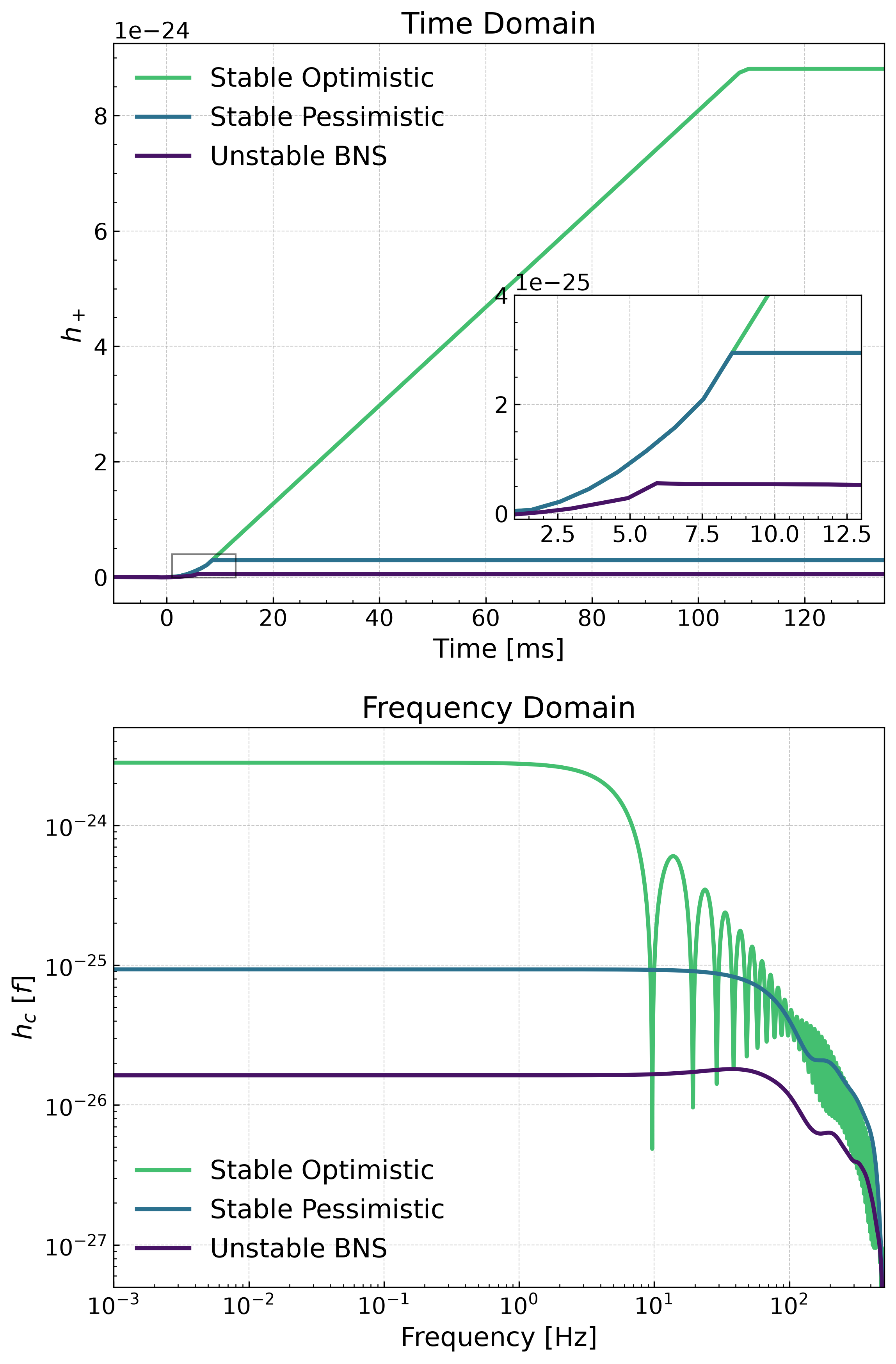}
    \caption{Neutrino memory signals from a BNS merger at $40$ Mpc with an inclination angle of $32^\circ$. Three scenarios are considered: stable remnant with optimistic (green) and pessimistic (blue) extrapolations, and unstable remnant that collapses into a black hole (purple).}
    \label{fig:neutrinos_results_all_cases}
\end{figure}

By applying Eq.~\ref{eq: h_plus_neutrinos} to the neutrino luminosities extracted from the simulations detailed in Sec.~\ref{subsub: simulations}, we evaluate the time evolution of the memory effect. The resulting strain amplitudes are presented in Fig.~\ref{fig:neutrinos_results_all_cases}, which contrasts the memory buildup for three post-merger possibilities.

\subsection{Gamma-ray burst}
Since both the prompt and afterglow phases involve a sudden acceleration of particles that subsequently become unbound, each phase can, in principle, be associated with its own distinct contribution to a memory effect.
\subsubsection{Prompt emission}

To estimate the memory effect of the prompt phase, we model the burst as a single point-particle of mass $m$ undergoing an acceleration from rest to a velocity $\beta$, over a timescale set by the prompt emission duration $T_{90}$. The point-particle is taken to carry the total prompt energy of the burst, and produces a memory amplitude $\Delta h_{\rm prompt}$ given by~\cite{Sago_2004,Birnholtz_2013}

\begin{equation}
    \Delta h_{\text{prompt}} = \frac{2 E_{\rm pp} \beta^2}{r},
    \label{eq: delta_h_grb}
\end{equation}

\noindent where $E_{\rm pp}$ is the point-particle energy, $\beta = v/c \simeq 1$ is the point-particle velocity and $r$ is the distance to the source. In this work, we set $E_{\rm pp} = E_{\gamma, \rm iso}$, i.e. we identify the fictitious mass with the radiated gamma photon energy budget rather than the kinetic energy of the jet~\cite{Akiba_2013} because we are interested in the prompt emission.

This model is also a simplified representation of the true jet dynamics, for two reasons. It neglects the finite acceleration timescale of the jet\footnote{Estimates suggest that the jet acceleration occurs on a timescale of order $0.1$ ms~\cite{Lopez:2023aja}, whereas in our model the rise time is instead assumed to be set by the prompt duration, and this shorter acceleration timescale is therefore neglected.}, treating the acceleration as instantaneous (neglecting the variability of the structure of the jet~\cite{Sago_2004}). Moreover, it reduces the jet to a point source rather than an object collimated within a finite half-opening angle $\theta_j$~\cite{Birnholtz_2013}. The assumption overestimates the true jet energy by a factor of $(1-\cos{\theta_j})$~\cite{Birnholtz_2013, Brabant:2026} as mentioned in Sec.~\ref{subsub: prompt_intro}, which scales as $\sim 2 \Gamma^2$ if the jet opening angle is comparable to the relativistic beaming cone. Thus, the model yields an optimistic upper bound on $\Delta h_{\rm prompt}$. For a typical opening angle of $30^{\circ}$, we overestimate the true jet energy by a factor of $\sim 7$. Nevertheless, we do not account for the opening angle impact further in this work as this parameter is generally not well constrained. The jet geometry is also neglected in this modeling. Following the top-hat and structured jet models of Birnholtz \& Piran (2013)~\cite{Birnholtz_2013}, the peak memory amplitude at a constant jet energy scales as $\theta_j^{-0.1}$ and $\theta_j^{-0.04}$, respectively~\cite{Birnholtz_2013}. Widening the opening angle also spreads the fixed energy over a larger solid angle, thereby reducing the maximum amplitude. The drop is steeper for the uniform top-hat jet whereas the structured jet mitigates the effect by keeping a high energy density in its narrow cone (of aperture $\sim \Gamma^{-1}$).

For an isotropic-equivalent energy of $E_{\gamma,\rm iso} = 10^{51}$ erg, our point-particle model gives the following order-of-magnitude estimate for the memory amplitude from a GRB prompt emission with $\Gamma = 100$ and a viewing angle of $0.1$ rad:

\begin{equation}
    \Delta h_{\text{prompt}} = 1.06 \times 10^{-24}
    \left(\frac{E_{\gamma,\rm iso}}{10^{51}~\text{erg}}\right)
    \left(\frac{r}{100~\text{Mpc}}\right)^{-1}.
\end{equation}
\subsubsection{Afterglow}

By analogy with the prompt GRB memory, the amplitude of the afterglow memory scales with the energy dissipated during this phase, characterized by its isotropic-equivalent value $E_{\rm K}~(\rm erg)$. The afterglow memory amplitude is then estimated, analogous to Eq.~\ref{eq: delta_h_grb}, as

\begin{equation}
    \Delta h_{\text{aft}} = \frac{2 E_{\rm K}\,\beta^2}{r},
    \label{eq: delta_h_aft}
\end{equation}

\noindent where $E_{\rm K}$ is the energy dissipated during the afterglow phase, $\beta = v/c$ is the velocity of the shocked material and $r$ is the distance to the source.

In our model, the relevant afterglow phase $T_{\rm aft}$ is defined as the interval between the deceleration time of the jet ($t_{\rm dec}$) and the jet break time ($t_{\rm break}$), such that $T_{\rm aft} \equiv t_{\rm break} - t_{\rm dec}$ (see Sec.~\ref{subsub: intro_aft} for more details).

\begin{figure}
    \centering
    \includegraphics[width=1\linewidth]{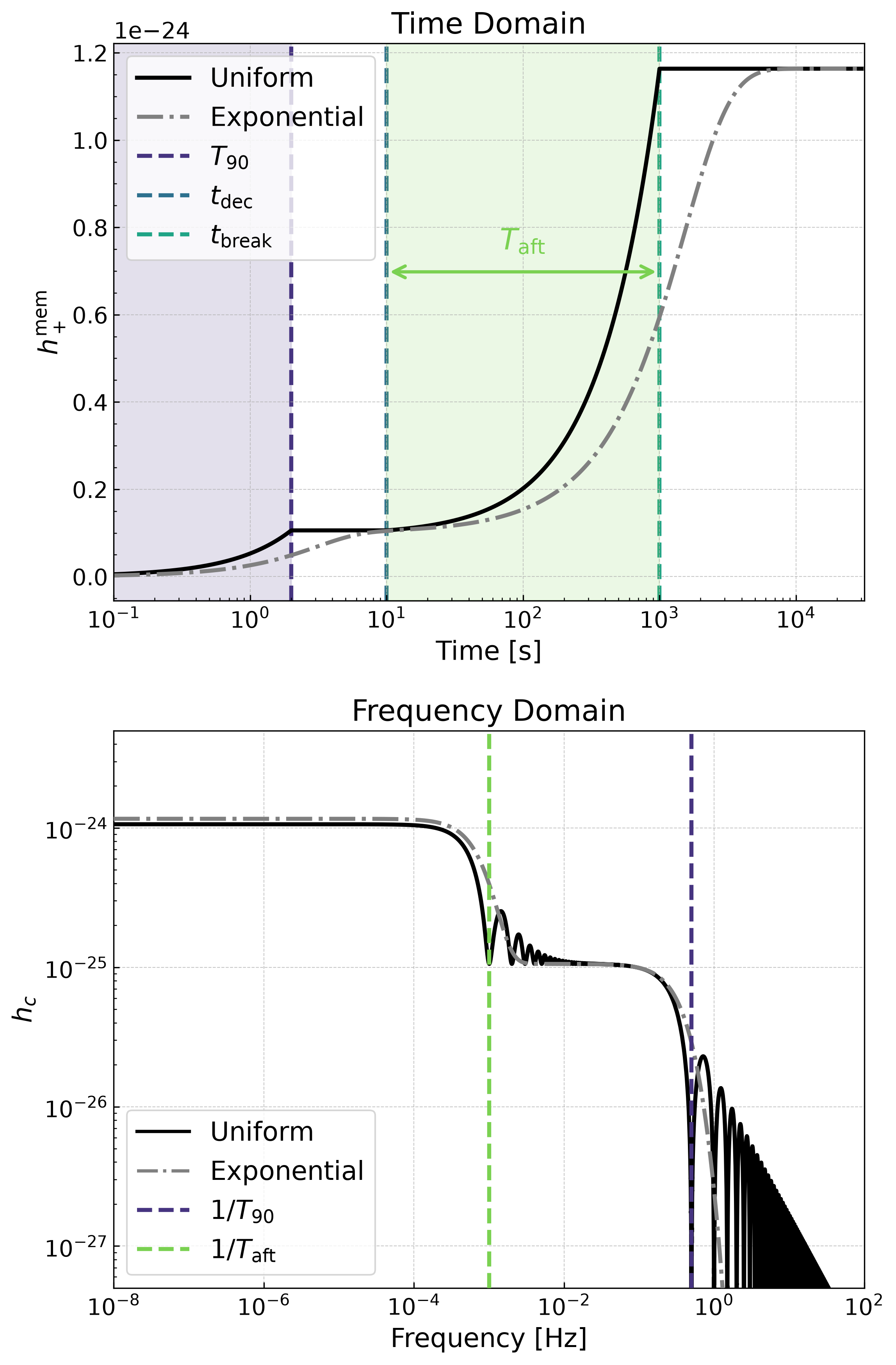}
    \caption{GRB memory signals in the time (\textit{top}) and frequency (\textit{bottom}) domains for a toy model with parameters: $E_{\rm prompt} = 10^{50}~{\rm erg}$, $E_{\rm aft} = 10^{51}~{\rm erg}$, $T_{\rm 90} =2~{\rm s}$, $t_{\rm dec} = 10~{\rm s}$, $t_{\rm break} = 1000~{\rm s}$, $\Gamma = 100$, $\theta = 0.1$ rad, and $r = 100~{\rm Mpc}$. Vertical lines refers to characteristic times/frequencies.}
    \label{fig:GRB_mem_toy_models_combined}
\end{figure}

In what follows, the modeling of the GRB memory will include both phases, as illustrated in Fig.~\ref{fig:GRB_mem_toy_models_combined} for a generic case. While our model remains simple, it is the first to date to consider these two phases together.
\subsection{Kilonova}

To evaluate the linear memory produced by the KN emission, we apply the exact same null-memory formalism used for the neutrino emission and replace the neutrino luminosity with the KN photon luminosity.

Consequently, the $h_+$ and $h_\times$ polarizations of the KN memory are computed using the expressions given in Eqs.~\ref{eq: h_plus_neutrinos} and~\ref{eq: h_cross_neutrinos}, simply substituting the neutrino luminosity distribution $dl_{\nu}/d\Omega'$ with the simulated kilonova luminosity distribution $dl_{\lambda}/d\Omega'$.

The energy and frequency of the emitted radiation are simulated for every point on the sky map at each time step. The time array covers a span starting $0.08$ days after the merger and ending $9.8$ days later, and is uniformly sampled at a frequency of $9.89 \times 10^{-4}\text{ Hz}$. The simulation reveals three distinct phases in the evolution of the KN luminosity, first a bright and strongly polar phase, then a fainter and more isotropic phase, and finally a late phase ($t \gtrsim 5$ days) where the polar and equatorial regions contribute comparably. The corresponding emission spectrum also evolves with time, shifting from short wavelengths ($ \sim 500$ nm) at early times to longer wavelengths ($ \sim 1500$ nm) at later times.

\begin{figure}
    \centering
    \includegraphics[width=1\linewidth]{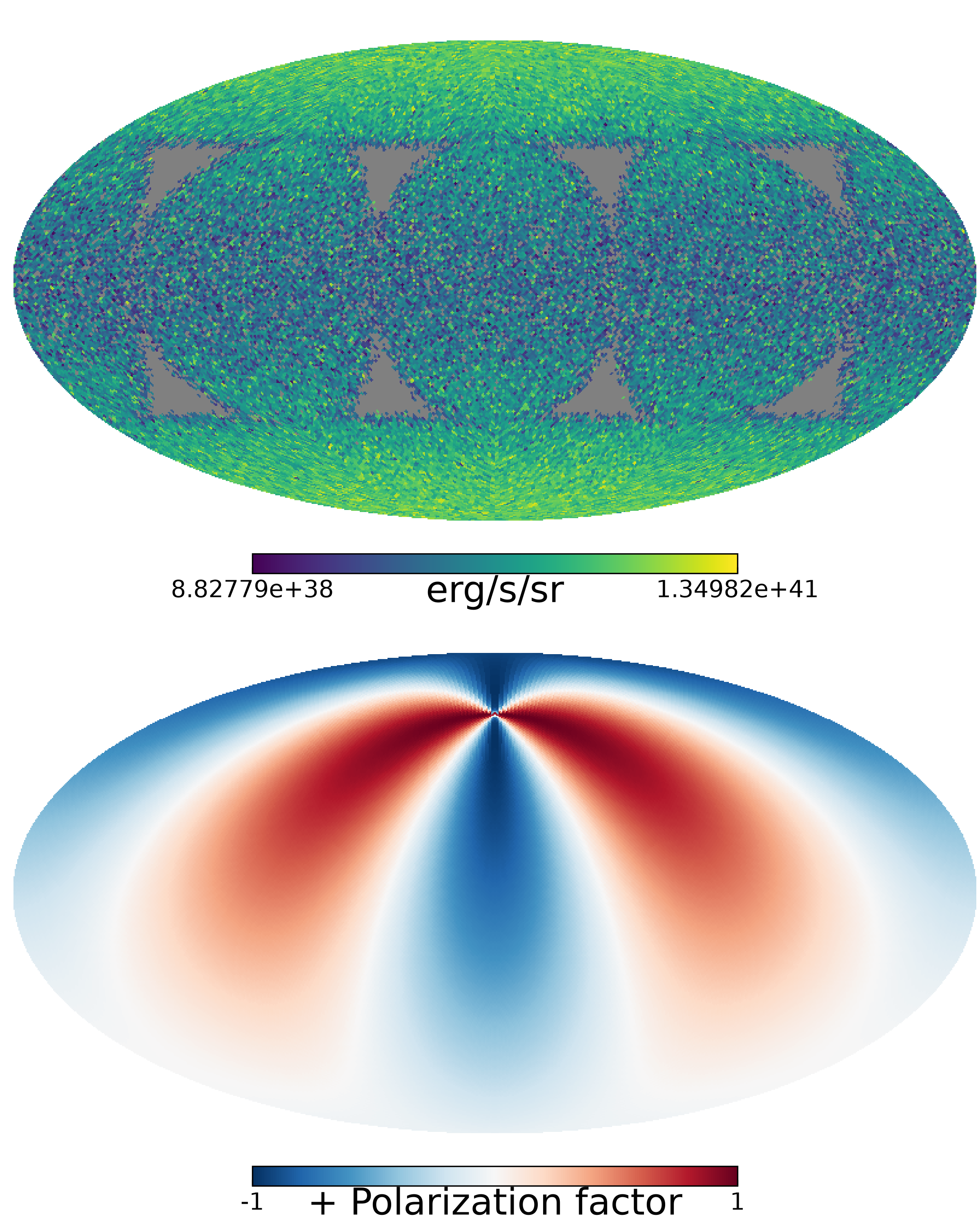}
    \caption{Comparison of the KN luminosity map 3 days after the merger (\textit{top}) and the angular factor (\textit{bottom}, for the + polarization) for a viewing angle $\theta = 32^\circ$ similar to GW170817 - AT2017gfo~\cite{Finstad_2018}. The grey patches in the luminosity map are data points with luminosity equal to zero.}
    \label{fig:comparison_maps}
\end{figure}

The memory signal is computed pixel by pixel from the simulated skymaps, by combining the local luminosity of each pixel, $dl_{\lambda}/d\Omega'$, with the angular factor $n^i n^j / (1-\cos\theta)$, projected onto the transverse-traceless basis\footnote{Here $\mathbf{z} =(0,0,1)$ is the unit vector pointing along the positive z-axis of a Cartesian reference coordinate system centered on the remnant. If $\mathbf{n}$ denotes the unit vector pointing from the source to the observer, we set $\mathbf{e}_1 = \mathbf{n} \times \mathbf{z}$. We also set $\mathbf{e}_2 = \mathbf{n}\times \mathbf{e}_1$. By construction, $\mathbf{e}_2$ is orthogonal to $\mathbf{n}$ and to $\mathbf{e}_1$, and forms with $\mathbf{e}_1$ an orthonormal basis in the transverse plane. We extract the polarizations by projecting $h^{\rm TT}$ and obtain:  
\begin{equation}
    h_+ = e_1^i e_1^j h_{ij}^{\rm TT}\ , \ \ \ h_\times = e_1^i e_2^j h_{ij}^{\rm TT} = 0 \rm \ \  \text{if } \phi' = 0.
\end{equation}}. The luminosity map therefore encodes the intrinsic brightness and geometry of the KN emission, while the angular factor encodes the geometric weighting associated with the observer's line of sight. A comparison of these two maps is shown in Fig.~\ref{fig:comparison_maps}, where the polar regions can be seen to dominate the luminosity map, while the angular factor map displays both positive and negative regions across the sky. The sign structure reflects the fact that the $+$ polarization can either shorten or lengthen the separation between test masses. As a result, an isotropic emission would lead to a cancellation between positive and negative contributions and the memory amplitude would be zero.

For a viewing angle consistent with the one inferred for GW170817~\cite{Finstad_2018}, the accumulated KN memory saturates at a final characteristic amplitude of order $1.6 \times 10^{-25}$.

In the frequency domain, the characteristic strain of the KN memory follows the same qualitative trend already found in the other cases, i.e. constant at low frequencies before decaying at higher frequencies. However, because the KN memory builds up over a much longer timescale, the transition occurs at correspondingly lower frequencies, around $10^{-4}$~Hz.

\section{Memory detectability in 3G detectors}
\label{sec:snr_and_case_study}
Having established the theoretical framework for both the nonlinear and linear memory contributions in Secs.~\ref{sec:nonlinear} and~\ref{sec:linear}, we present the SNR formalism used for ET and LISA in order to estimate the detectability of the memory effect.

\subsection{SNR in Einstein Telescope}
We quantify how the memory signal stands out from the noise by computing the SNR, $\rho$. Unlike conventional GW searches, which rely on matched filtering against a template bank of waveforms to extract a known signal from the noise, the memory signal is similar to an unmodeled burst. Its detectability is instead assessed directly from its frequency content. The key quantity is then the characteristic strain $h_c(f) = 2 f|\tilde{h}|$, which is compared to the detector noise amplitude spectral density over the relevant sensitive band. The SNR is then defined as~\cite{Favata_2010_intro, Flanagan:1997kp}

\begin{equation}
    \rho = \left[ \int_{f_{\rm min}}^{f_{\rm{max}}} \rm{d\ln{f}}~ \frac{h_c^2 (f)}{h_n^2 (f)}\right]^{1/2} ,
    \label{eq:snr_def}
\end{equation}
\noindent where $h_n$ is the noise amplitude for a detector with noise power spectral density $S_n (f)$, defined as $h_n = \sqrt{f S_n(f)}$. 

The detectability of the signal is visually assessed by comparing the characteristic strain with the detector sensitivity curve. A signal whose characteristic strain lies above the sensitivity curve will accumulate a higher SNR and thus be more readily detectable.

For ET, Eq.~\ref{eq:snr_def} is evaluated using the ET-D noise spectral density configuration~\cite{Hild_2011, ET_sensitivity_curve}, with the integral computed over ET's sensitive band, $f_{\rm min} = 1$~Hz and $f_{\rm max} = 10^{4}$~Hz. To claim a detection, we adopt a conservative threshold of $\rho > 8$~\cite{Abadie:2010}.

To account for the multimessenger nature of the merger, the total SNR of an event consisting of $N$ different counterparts is estimated by summing in quadrature the individual $\rho_i$~\cite{Lopez:2023aja}
\begin{equation}
    \rho_{\rm tot} = \sqrt{\sum_{i=1}^N \rho^2_i }.
    \label{eq: SNR_tot_ET}
\end{equation}

\subsection{SNR in LISA}
For LISA, tools are available to simulate a GW signal onto the detector arms and to estimate the interferometer's response more accurately than with the characteristic strain. In this approach, the positions of the satellites forming the interferometer and the response of the arms must be provided. The satellite orbits are generated using the \verb|lisaorbits| package~\cite{LISA_Orbits} over a four-year duration adopting the \verb|EqualArmlengthOrbits| configuration (Keplerian orbits that minimize flexing to leading order in eccentricity). The memory signal is then simulated through the satellite constellation using \verb|LISAGWResponse|~\cite{LISA_GW_Response}, which returns the six single-link response time series. Each of these time series encodes the relative Doppler shift induced along one of the six one-way laser links between the three LISA spacecraft, as perturbed by the passage of a GW~\cite{Inchauspe:LISA}. These link responses are then combined to simulate the full interferometer output. In our case, we work with the three nearly independent second-generation Time-Delay Interferometry (TDI) channels, A, E, and T, obtained from the six link responses using \verb|pytdi| pipeline~\cite{pytdi}. A more detailed description of the TDI method as well as an illustration is shown in Fig.~2 of \cite{Inchauspe:LISA}.

Finally, the SNR $\rho_C$ ($C \in \{A, E, T\}$) for each channel is~\cite{Inchauspe:LISA}
\begin{equation}
    \rho^2_C = 4~\Re \int_{f_{\rm min}}^{f_{\rm max}} \frac{\tilde{d}(f)\tilde{d}^*(f)}{S_{CC}}~df,
\end{equation}
where $\tilde{d}(f)$ is the FT of the TDI data in channel $C$, the superscript $*$ denotes complex conjugation, and $S_{CC}(f)$ is the one-sided power spectral density of the corresponding channel's noise. The integration bounds are set by LISA's sensitive band, $f_{\rm min} = 5\times10^{-5}$~Hz and $f_{\rm max} = 1$~Hz. The total SNR across the three channels is obtained by summing the individual channel contributions in quadrature~\cite{Inchauspe:LISA},
\begin{equation}
    \rho_{\rm tot} = \left( \rho_A^2 + \rho_E^2 + \rho_T^2 \right)^{1/2},
    \label{eq:SNR_tot_LISA}
\end{equation}
\noindent consistent with the definition of Eq.~\ref{eq: SNR_tot_ET}.

\section{A case study: GW170817 - GRB 170817A - AT2017gfo}
\label{sec:170817_case}
In this section, we apply the formalism developed above to the specific case of the BNS observed on 17 August 2017, computing the memory effects discussed in Sec.~\ref{sec:nonlinear} and Sec.~\ref{sec:linear} with the method in Sec.~\ref{sec:snr_and_case_study}.

\subsection{Mapping the parameter space}
\begin{figure}

    \includegraphics[width=1\linewidth]{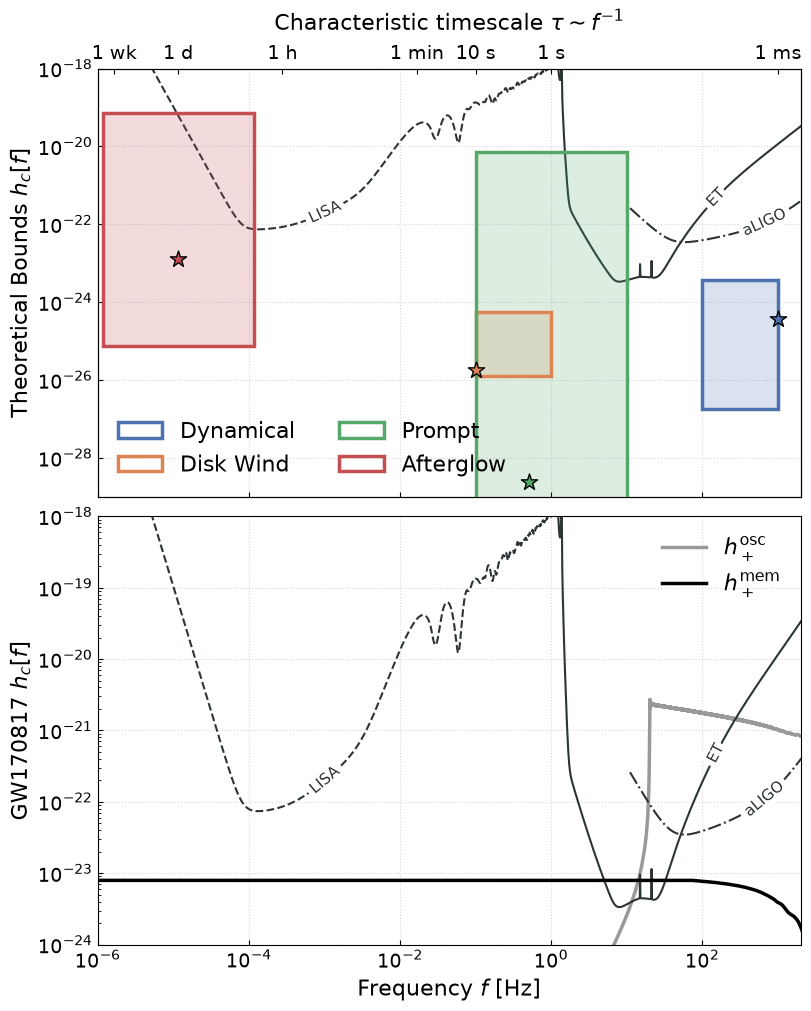}

    \caption{Summary of the characteristic strain $h_c(f)$ for a BNS merger at a distance similar to GW170817. \textit{Top:} Plausible parameter ranges for the linear memory effect bounded by the parameter values in Table~\ref{tab:min_max_parameters} using our toy-model. The stars denote the estimates for GW170817 used in this work and listed in Table~\ref{tab:gw170817_params}. \textit{Bottom:} Characteristic strain of the oscillatory waveform $h^{\mathrm{osc}}_+$ (grey curve) and the nonlinear memory $h^{\mathrm{mem}}_+$ (black curve) for GW170817. The oscillatory component is computed using the hybridized numerical relativity approximant NRHybSur3dq8~\cite{Varma_2019}.}
    \label{fig:hc_vs_f_LISA_ET}
\end{figure}

Before analyzing the specific memory effects from GW170817, we first map the theoretical boundaries of the parameter space to identify which counterparts are most likely to produce detectable memory signals. Thanks to our models described in Sec.~\ref{subsub: toy_model}, we illustrate the sensitive regions of the two detectors (ET and LISA) for three linear memory signals: dynamical ejecta, disk wind ejecta, and GRB. For these counterparts, plausible parameter ranges can be estimated either from theoretical prediction (dynamical and disk-wind ejecta) or from observations (GRB). The corresponding minimum and maximum values used to bound each region are listed in Table~\ref{tab:min_max_parameters}. Based on these ranges, Fig.~\ref{fig:hc_vs_f_LISA_ET} demonstrates the possible detection regions for each counterpart, based only on their characteristic strain for a system at $40$~Mpc derived from our toy model. The bottom panel is meant to illustrate the characteristic strain profile of the oscillating and (nonlinear) memory components of a waveform similar to GW170817.

\begin{table}
\caption{Parameter ranges used to estimate the minimum and maximum
characteristic strain for the dynamical, disk wind ejecta~\cite{Metzger:2019zeh, Nakar:2019fza}, and GRB counterpart~\cite{fermi_duration_distrib, Laskar_2023, Rhoads_1999, Sari:break, Fong_2015}.} 
\label{tab:min_max_parameters}
\begin{ruledtabular}
\begin{tabular}{llcc}
Component & Parameter & Min & Max \\
\midrule
\multirow{2}{*}{Dynamical ejecta} 
& $\log_{10}(\text{Mass})~[M_\odot]$ & $-4$ & $-2$ \\
& Velocity [$c$]        & $0.1$     & $0.4$\\     
& $\tau_{\rm dyn}$~[ms] & $1$ & $10$\\
\midrule
\multirow{3}{*}{Disk wind ejecta} 
& $\log_{10}(\text{Mass})~[M_\odot]$ & $-3$ & $-2$ \\
& Velocity [$c$]         & $0.1$     & $0.2$\\ 
& $\tau_{\rm wind}$~[s]  & $1$ & $10$  \\
\midrule
\multirow{2}{*}{GRB prompt}        
& $\log_{10}(E_{\gamma,\rm{iso}})~[\mathrm{erg}]$ & $46$ & $55$ \\ 
& $T_{90}$ [s] &  $0.1$ & $10$  \\
\midrule
\multirow{2}{*}{GRB afterglow}     
& $\log_{10}(E_{\rm{K}})~[\mathrm{erg}]$ & $50$ & $56$ \\
& $T_{\rm aft}$ [days]  & $0.1$     & $10$      \\
\end{tabular}
\end{ruledtabular}
\end{table}

The dynamical and disk-wind ejecta fall below the ET sensitivity curve. For the GRB, we distinguish the prompt and afterglow phases. The short-duration prompt emission is more likely to be detectable with ET, whereas the very long-duration (more than $100$~s) afterglow emission would instead approach LISA high-frequency range. For the afterglow, only extremely short durations (between $0.1$ and $1$~day) yield a characteristic strain above LISA sensitivity curve. \ignore{However, the TDI projection typically reduces the effective amplitude by about two orders of magnitude compared to the characteristic strain, making the detection of the memory signal from short afterglows difficult to achieve.} 

\subsection{Time-domain signal}
To see how these theoretical bounds translate into a more realistic multimessenger scenario, we now apply our comprehensive modeling to the event of August 2017. We use the parameters listed in Table~\ref{tab:gw170817_params} to compute the nonlinear memory and the linear memory induced by the dynamical ejecta, disk wind ejecta, GRB. The KN and neutrino memory are instead estimated directly from the simulations presented in Sec.~\ref{subsub: simulations}, that have been rescaled to a distance of $40$ Mpc. 

\begin{table}[!ht]
\centering
\caption{Parameters used for the computation of the memory effect from a BNS system similar to GW170817 - GRB 170817A and ejecta from AT2017gfo. The memory waveform is computed using NRSur7dq2 approximant as implemented in \texttt{GWMemory}.}

\begin{ruledtabular}
\begin{tabular}{l c}
Component & Value \\
\midrule
\multicolumn{2}{l}{\textbf{Binary system}~\cite{discovery_170817, Finstad_2018}} \\
Primary mass $m_1$ [$M_\odot$] & $1.46$ \\
Secondary mass $m_2$ [$M_\odot$] & $1.27$ \\
Mass ratio $q$ & $1.185$ \\
Total mass $M$ [$M_\odot$] & $2.73$ \\
Luminosity distance $r$ [Mpc] & $40$ \\
Inclination angle $\theta$ [$^\circ$] & $32$ \\
\midrule
\multicolumn{2}{l}{\textbf{Dynamical ejecta}~\cite{KN_estim_170817, Kasen:2017sxr}} \\
Ejecta mass $M_{\rm ej,dyn}$ [$M_\odot$] & $0.005$ \\
Ejecta velocity $v_{\rm ej,dyn}$ [$c$] & $0.25$ \\
Characteristic timescale $\tau_{\rm dyn}$ [s] & $10^{-3}$ \\
\midrule
\multicolumn{2}{l}{\textbf{Wind ejecta}~\cite{AT2017gfo_discovery, Siegel:2017nub, Villar_2017}} \\
Ejecta mass $M_{\rm ej,wind}$ [$M_\odot$] & $0.01$ \\
Ejecta velocity $v_{\rm ej,wind}$ [$c$] & $0.05$ \\
Characteristic timescale $\tau_{\rm wind}$ [s] & $10.0$ \\
\midrule
\multicolumn{2}{l}{\textbf{GRB prompt}~\cite{GRB_detect_170817}} \\
Isotropic energy $E_{\gamma,\rm iso}$ [erg] & $3\times10^{46}$ \\
Prompt duration $T_{90}$ [s] & $2.0$ \\
\midrule
\multicolumn{2}{l}{\textbf{GRB afterglow}~\cite{Li_2021}} \\
Isotropic energy $E_{\rm K}$ [erg] & $10^{52.2}$ \\
Deceleration time $t_{\rm dec}$ [s] & $50$ \\
Break time $t_{\rm break}$ [day] & $1$ \\
\end{tabular}
\end{ruledtabular}
\label{tab:gw170817_params}
\end{table}

\begin{figure}
    \centering

    \centering
    \includegraphics[width=1\linewidth]{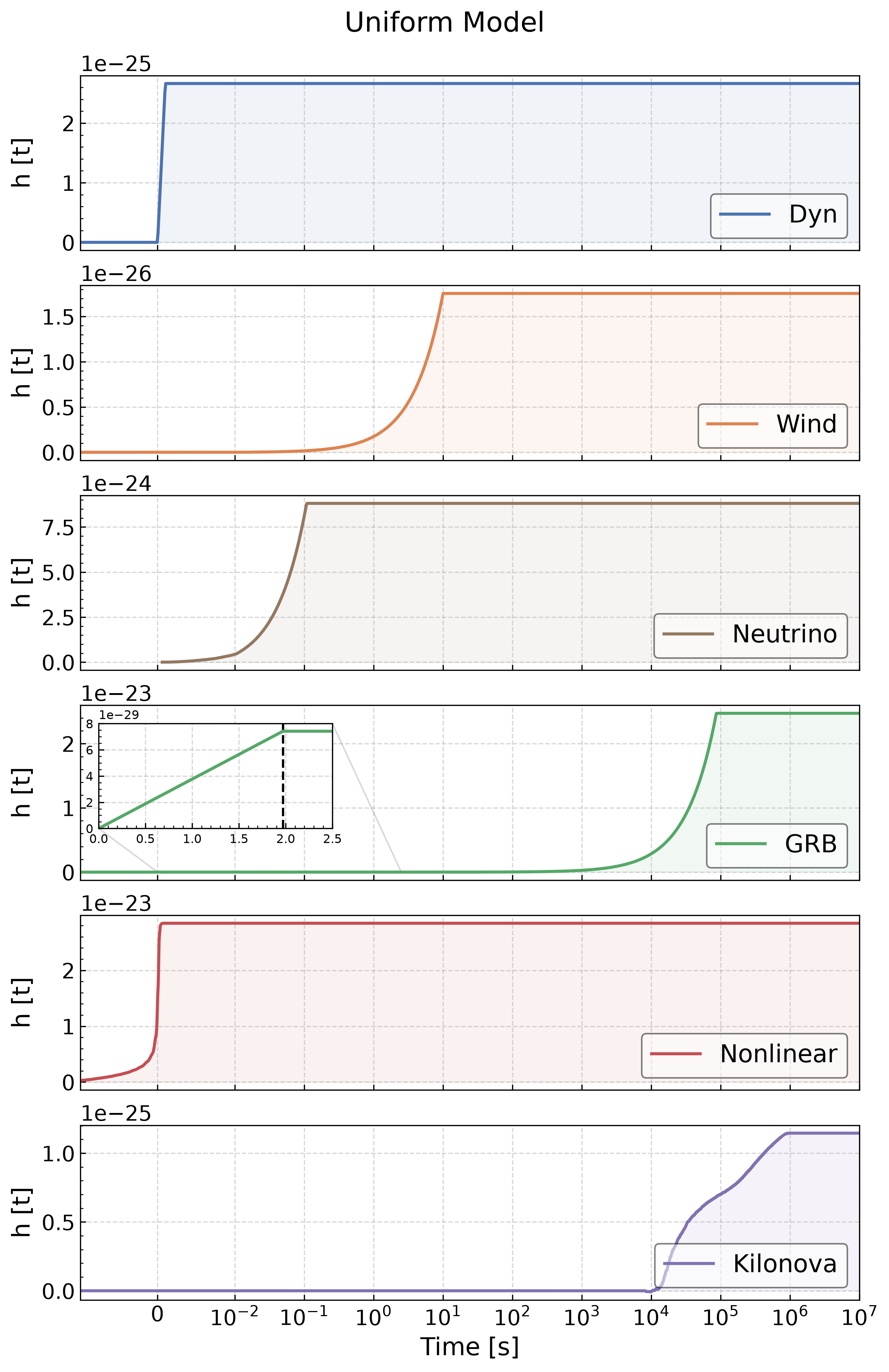}
    \caption{Timeseries of the different linear and nonlinear memory signatures in a BNS merger similar to GW170817 with the uniform model for the dynamical, disk wind and GRB memory.  The dashed vertical line in the inset refers to $t = T_{\rm90}$. For the neutrino counterpart, only the optimistic scenario is displayed.}
    \label{fig:ts_170817}
\end{figure}

Fig.~\ref{fig:ts_170817} displays the resulting time evolution of the linear and nonlinear memory contributions across the various BNS signatures. The amplitude from the nonlinear memory is dominant, followed by the GRB emission and the (optimistic) neutrino counterpart. The prompt emission produces a much weaker memory, of order $10^{-29}$ as shown in the inset, due to the low prompt energy of GRB 170817A. The disk wind ejecta lies one order of magnitude below the dynamical ejecta, since it is a lower-velocity component and the memory amplitude scales quadratically with the velocity. Finally, the KN reaches an amplitude comparable to the dynamical ejecta, but evolves over a much longer timescales.  

\subsection{Prospects for Einstein Telescope}
\begin{figure}
    \centering

        \includegraphics[width = 1.\linewidth]{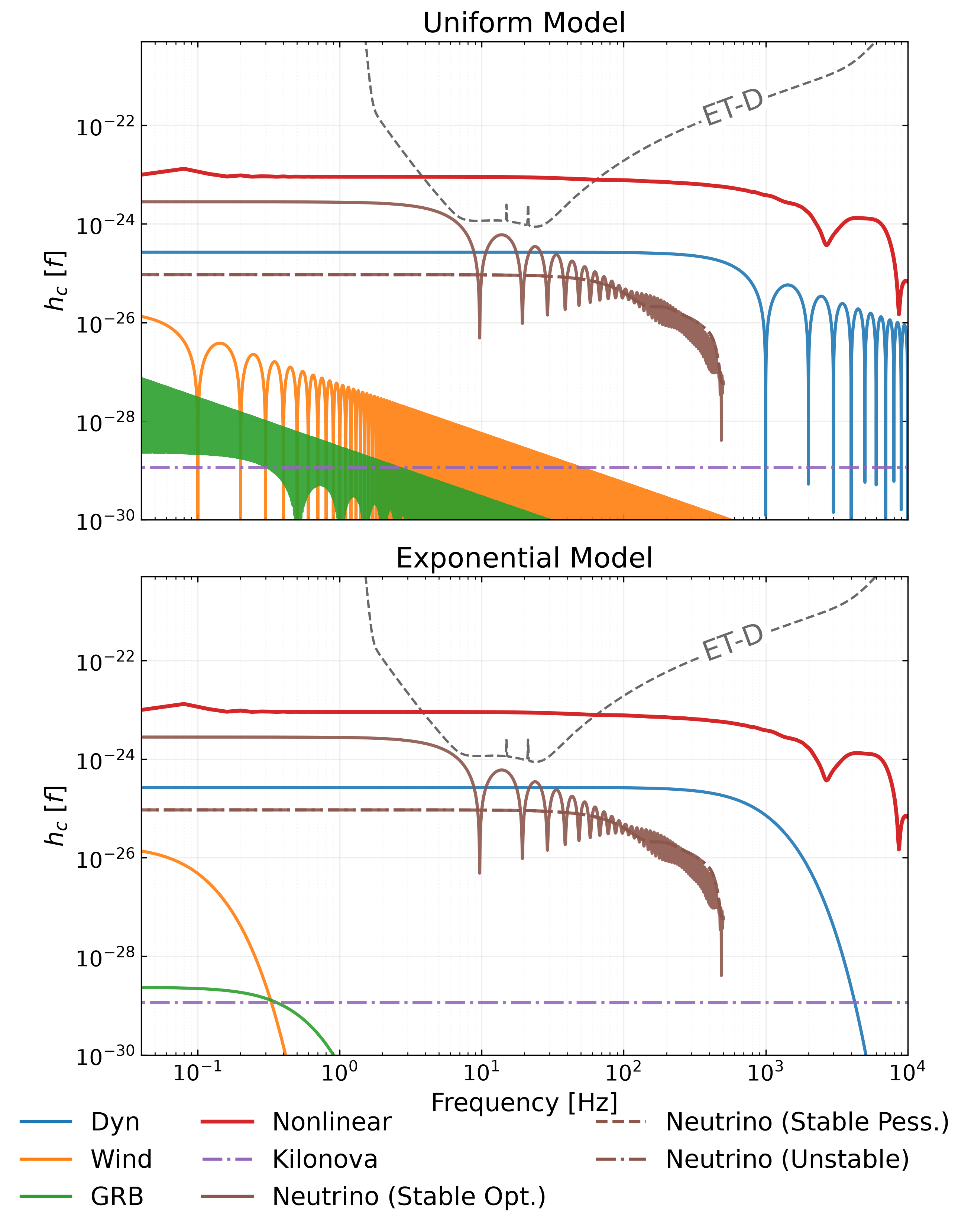}
        \caption{Comparison of the uniform (\textit{top}) and exponential (\textit{bottom}) models for the characteristic strains of an event similar to GW170817. The KN strain has been extrapolated up to the sensitivity range of ET.}
        \label{fig:170817_comparison}

\end{figure}

Fig.~\ref{fig:170817_comparison} shows the characteristic strain of each memory component in ET, for both the uniform and exponential rising-phase models. Under the exponential model, the disk wind and GRB curves decay much faster for frequencies above $\tau^{-1}$. In addition, the decay phase of these two components fall within the sensitivity curve of ET. The same behavior is even more pronounced for the KN, whose decay starts close to $10^{-5}$~Hz. The signal was therefore extrapolated up to $10^{4}$~Hz. The dynamical ejecta, by contrast, shows a comparable strain under both models, owing to its much shorter characteristic timescale.

Notably, among the linear memory components, the GRB (prompt and afterglow) strain is strongly suppressed at high frequencies. However, its behaviour at lower frequencies (than ET sensitive band) is not shown and could present a higher amplitude. The reasons for this are related to the afterglow component, which has a longer characteristic time (and thus a low-frequency component) and higher energy than the weak prompt emission. Although the behavior of the disk wind curve is similar to that of the GRB, the amplitude range estimated in the previous sections does not indicate improved detectability at lower frequencies.

To quantify the detectability, the SNR is computed from Eq.~\ref{eq:snr_def} using the parameters listed in Table~\ref{tab:gw170817_params}. The resulting SNR values are summarized in Table~\ref{tab:snr_component_models}.

\begin{table}
\centering
\caption{SNR contributions per memory component for GW170817-GRB 170817A-AT2017gfo case in ET. For components that depend on the step function model (dynamical ejecta, wind, GRB), results are shown for both the uniform and exponential models.}
\label{tab:snr_component_models}
\begin{ruledtabular}
\begin{tabular}{l c}
Component & $\rho$ \\
\midrule
\textbf{Dynamical ejecta} \\
\quad Uniform & $0.3248$ \\
\quad Exponential & $0.3248$ \\
\midrule
\textbf{Disk wind ejecta} \\
\quad Uniform & $3.995 \times 10^{-5}$ \\
\quad Exponential & $< 10^{-9}$ \\
\midrule
\textbf{GRB} \\
\quad Uniform & $2.108 \times 10^{-6}$ \\
\quad Exponential & $< 10^{-9}$ \\
\midrule
\textbf{Kilonova} & $1.392 \times 10^{-5}$ \\
\midrule
\textbf{Nonlinear} & $10.768$ \\
\midrule
\textbf{Neutrino scenarios} \\
\quad Stable optimistic & $0.6091$ \\
\quad Stable pessimistic & $0.1101$ \\
\quad Unstable & $2.087 \times 10^{-2}$ \\
\midrule
\textbf{Total SNR} \\
\quad Without neutrinos & $10.773$ \\
\quad With neutrinos (optimistic) & $10.790$ \\
\quad With neutrinos (pessimistic) & $10.778$ \\
\quad With neutrinos (unstable) & $10.773$ \\
\end{tabular}
\end{ruledtabular}
\raggedright \footnotesize \textbf{Note:} Total SNRs are quasi identical for both models because it is dominated by the nonlinear memory.
\end{table}

The very low SNR values for the disk wind and GRB components under the exponential model directly reflect the steeper high-frequency decay of their characteristic strain noted above. The dynamical ejecta, whose characteristic timescale is much shorter, is comparatively unaffected by this choice of model. Hence, the exponential model strongly penalizes counterparts whose $\tau^{-1}$ lies below the sensitive band of the detector. 

If we exclude the dominant nonlinear component, the dynamical ejecta and neutrinos are the primary linear sources. For the dynamical ejecta, its characteristic time is optimal for detection with ET. Its amplitude, although too weak for a detection, is overestimated due to the simplified modeling with the point-particle assumption. For neutrinos, the case of a stable remnant, when extrapolated optimistically, yields an SNR nearly double that of the dynamical ejecta, whereas the pessimistic case is close to $\rho \sim 0.1$. The discrepancy between these two neutrino scenarios implies that the dynamical ejecta and the possible (undetected) neutrino flux from GW170817 would have produced a nearly identical SNR in ET.

Overall, the total SNR associated with 170817 is set by the nonlinear memory contribution and is equal to $10.7$, high enough to be considered detectable.

\subsection{Prospects for LISA}
\begin{figure}
    \centering
    \begin{minipage}{\columnwidth}
    \centering
    \includegraphics[width=1\linewidth]{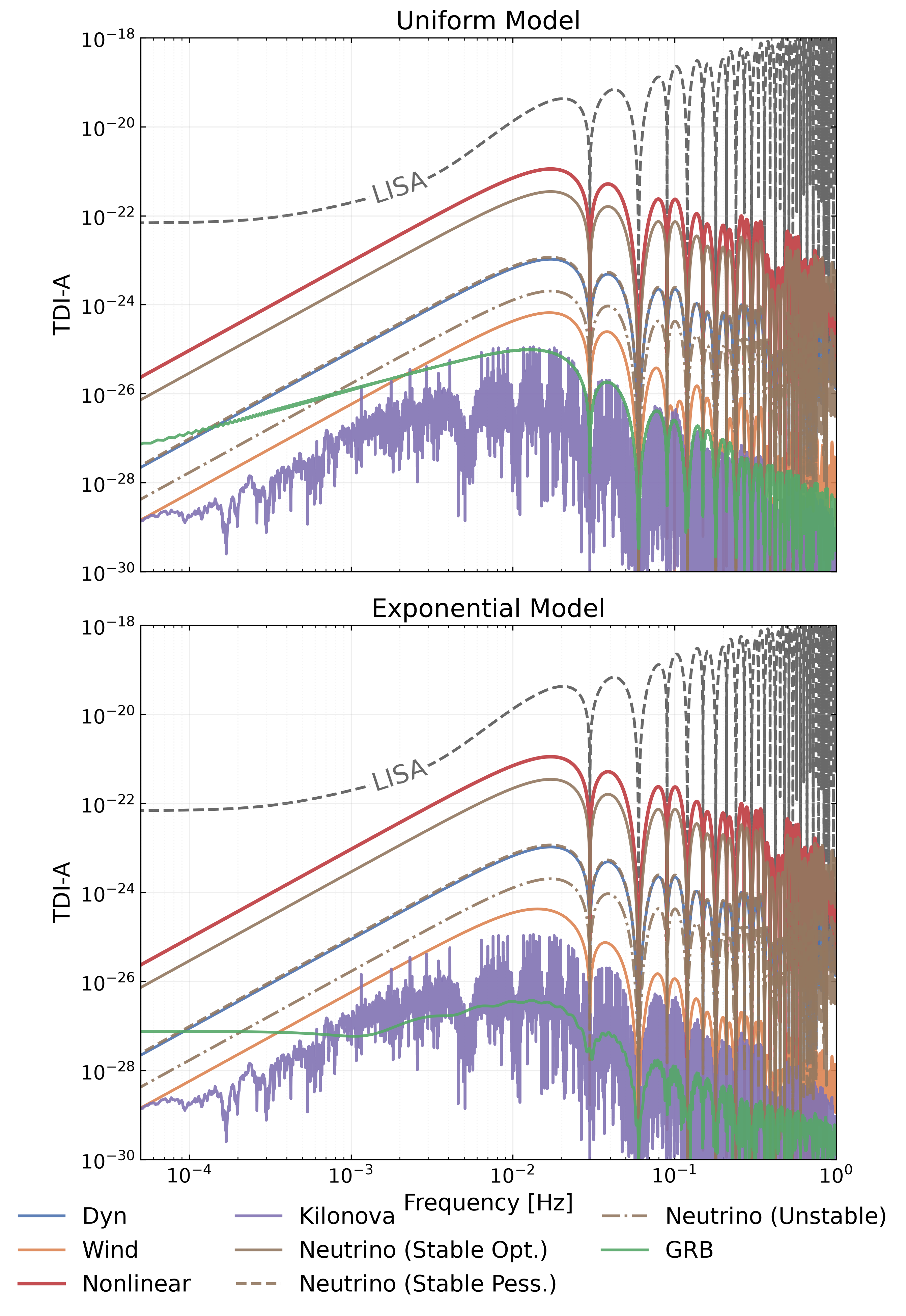}

    \caption{Comparison of the uniform (\textit{top}) and exponential (\textit{bottom}) models for the TDI-A outputs of all the counterpart studied.}
    \label{fig:tdi_memory_all_components}
    \end{minipage}
\end{figure}

For low frequencies covered by LISA, Fig.~\ref{fig:tdi_memory_all_components} shows the same memory signatures projected through the LISA response and combined here into the TDI-A channel. The relative hierarchy between components remains unchanged. Among the three components modeled with the two step-function models, only the GRB shows a noticeably smoother low-frequency rise under the exponential model. This difference arises because GRB is a two-phase event, characterized by two timescales: the prompt emission and the afterglow. The latter, in particular, is expected to affect low frequencies, which explains why the models diverge around $10^{-3}$~Hz. The dynamical and disk-wind ejecta by contrast remain quasi-identical since their characteristic timescales are shorter than $10$~s and thus only affect the higher and negligible frequencies within LISA band.

The SNR estimates reported in Table~\ref{tab:snr_component_models_lisa} confirm that the signals are too weak to result in a detection in LISA ($\rho_{\rm tot} \lesssim 2\times 10^{-2}$). 

\begin{table}
\centering
\caption{SNR contributions per memory component for LISA.}
\label{tab:snr_component_models_lisa}
\begin{ruledtabular}
\begin{tabular}{lc}
Component & $\rho$ \\
\midrule
\textbf{Dynamical ejecta} \\
\quad Uniform & $2.2 \times 10^{-4}$ \\
\quad Exponential & $2.2 \times 10^{-4}$ \\
\midrule
\textbf{Disk wind ejecta} \\
\quad Uniform & $1.2 \times 10^{-5}$ \\
\quad Exponential & $1.4 \times 10^{-5}$ \\
\midrule
\textbf{GRB} \\
\quad Uniform & $7.6 \times 10^{-6}$ \\
\quad Exponential & $3.2 \times 10^{-7}$ \\
\midrule
\textbf{Kilonova} & $1.1 \times 10^{-6}$ \\
\midrule
\textbf{Nonlinear} & $2.3 \times 10^{-2}$ \\
\midrule
\textbf{Neutrino scenarios} \\
\quad Stable optimistic & $7.2 \times 10^{-3}$ \\
\quad Stable pessimistic & $2.4 \times 10^{-4}$ \\
\quad Unstable & $4.2 \times 10^{-5}$ \\
\midrule
\textbf{Total SNR} \\
\quad Without neutrinos & $1.87 \times 10^{-2}$ \\
\quad With neutrinos (optimistic) & $1.96 \times 10^{-2}$ \\
\quad With neutrinos (pessimistic) & $1.87 \times 10^{-2}$ \\
\quad With neutrinos (unstable) & $1.87 \times 10^{-2}$ \\
\end{tabular}
\end{ruledtabular}
\end{table}

Even though the GRB afterglow has a characteristic duration of the order of a day, which suggests that its memory signal should fall within LISA optimal band, its SNR contribution remains low. This comes from the fact that the amplitude of the memory is modeled via the total available energy, not by the timescale alone. For the underluminous GRB 170817A, the energy budget is too low to overcome LISA noise, regardless of how well the frequency of the signal aligns with the detector sensitivity band.

\section{Focus on Gamma-ray burst memory effect}
\label{sec:GRBs}
In the comparison of the memory channels presented above, the GRB signal is the one most likely to have an impact at the low frequencies covered by LISA. Its characteristic strain can take large values outside the sensitivity range of ET (see Sec.~\ref{sec:170817_case}), and it appears to be more variable (in terms of modeling) in LISA TDI channels. In addition, GRB 170817A is itself an atypical, sub-luminous event, so its own memory signal cannot alone establish whether GRB memory is a promising source class for 3G detectors. Recent and upcoming wide-field surveys are expected to increase the rate of detected GRBs, potentially revealing a population with properties that differ from those observed to date, and for which no well-established population model currently exists. 

To assess whether GRBs are a source of memory effects of interest in future detectors, we propose a second study focused on this counterpart. We proceed in two steps. We first select a set of observed GRBs that covers the extremes of the relevant parameter space for both the prompt and afterglow part. We then use these bounds to generate two synthetic GRB population. A first one with flat prior on all parameters, to determine if any GRBs could produce detectable a memory effect. The second one involves weighting the observed parameters to obtain a more physically motivated distribution. 

\subsection{Observed gamma-ray bursts}

To bracket the parameter space populated by real events, we select four additional GRBs that represent extreme cases along the dimensions relevant to the memory amplitude together with GRB~170817A itself. The BOAT~\cite{Burns_2023} (``Brightest Of All Time'', GRB~221009A) is the most energetic prompt emission ever recorded, with $E_{\mathrm{\gamma,iso}} = 10^{55}$~erg~\cite{Burns_2023}, and thus sets the high-energy boundary of the distribution. GRB~250702B is, to date, the longest GRB observed~\cite{Neights_2025}, with a prompt phase extending over three distinct emission episodes and a total duration $T_{90} \sim 2.5\times10^4$~s~\cite{Neights_2025}. The latter sets the upper bound on burst duration and is also one of the most distant events considered here. GRB~980425~\cite{1998Natur.395..670G}, associated with the supernova SN~1998bw~\cite{1999ApJ...516..788W}, is the closest GRB ever detected ($r \simeq 37$~Mpc~\cite{1998Natur.395..670G}) and anchors the low end of the energy range, being sub-luminous ($E_{\mathrm{\gamma,iso}} \sim 6\times10^{47}$~erg~\cite{1998Natur.395..670G}). Finally, GRB~211211A~\cite{Rastinejad_2022} is included as a second KN-associated GRB, analogous to GRB~170817A, but for which no GW counterpart was identified. This case therefore probes the memory signal expected from a KN-associated GRB without the additional constraint of a GW detection. Together with GRB~170817A, these five events span roughly nine orders of magnitude in prompt energy, a factor of $\sim 200$ in distance, and four orders of magnitude in duration, and their properties are summarized in Table~\ref{tab:grb_parameters_portrait}.

\begin{table}
\caption{Properties of selected GRBs}
\label{tab:grb_parameters_portrait}
\begin{ruledtabular}
\begin{tabular}{lccccc}
GRB & $r$ & $E_{\gamma,\rm{iso}}$ & $E_{\rm{K}}$ & $T_{90}$  \\
    & [Mpc] & [erg] & [erg] & [s]  \\
\colrule
BOAT~\cite{Burns_2023, Laskar_2023}        & 724    & $1 \times 10^{55}$   & $3 \times 10^{54}$ & 600           \\
250702B~\cite{Neights_2025}     & 7093   & $2.2 \times 10^{54}$ & $8 \times 10^{54}$ & $25\,000$   \\
980425~\cite{1998Natur.395..670G}      & 36.9   & $6 \times 10^{47}$   & $1 \times 10^{49}$ & 30           \\
170817A~\cite{GRB_detect_170817}     & 40     & $3 \times 10^{46}$   & $10^{52.2}$        & 2              \\
211211A~\cite{Rastinejad_2022}     & 350    & $7.6 \times 10^{51}$ & $5 \times 10^{52}$ & 51.37       \\
\end{tabular}
\end{ruledtabular}
\begin{minipage}{\columnwidth}
\vspace{2pt}
\footnotesize
\raggedright
\noindent\textbf{Notes:} The distance of GRB 250702B has been converted from redshift ($z = 1.036$) using $H_0 = 67.66$~km~s$^{-1}$~Mpc$^{-1}$. For all GRBs, we set $\Gamma = 100$ and the afterglow duration $T_{\rm aft}$ to $1$~day.
\end{minipage}
\end{table}

\subsection{Gamma-ray burst simulated population}

We then generate a synthetic population of GRBs and evaluate the memory SNR across the resulting parameter space. However, numerous GRBs of different origin (BNS merger or collapsar scenarios) have been detected, and no well-established model currently reproduces the full observed distribution of GRB properties.

Given the lack of a common GRB population model, we construct an empirical population based on the observed extreme cases discussed above. Each range of parameter values is set by the values spanned by the five events of Table~\ref{tab:grb_parameters_portrait}, added to the bimodal short/long prompt-duration distribution of the Fermi/GBM catalog~\cite{fermi_duration_distrib}, summarized in Table~\ref{tab: optimistic_population}. All timescales are assumed to be in the observer frame, so no explicit redshift correction is applied. Within this common parameter space, we consider two sampling strategies, which share the same parameter ranges but differ in how these ranges are populated.

In the first approach, parameter values are uniformly sampled within their respective ranges using a Sobol Quasi-Monte Carlo (QMC)~\cite{SOBOL196786} from the SciPy library~\cite{2020SciPy-NMeth}, that produce a sequence with an even coverage of the eight-dimensional space. This choice makes no assumption about the relative likelihood of different parameter values, and as a result tends to produce more extreme joint configurations (e.g. simultaneously high energy and small distance) than a more realistic population would. 

In the second approach, the same parameter ranges are instead populated according to the observationally motivated\footnote{And probably not accurate,  given the discussion on the current existence of no global population models.} prior distributions displayed in Table~\ref{tab: optimistic_population}, sampled via a Sobol QMC sequence. These priors assign a probability weight to extreme parameter values, consistent with the fact that, e.g., highly energetic and very nearby GRBs are intrinsically rarer than moderate ones. This yields a more realistic, if still approximate, representation of the underlying GRB population, and correspondingly suppresses the occurrence of extreme joint configurations relative to the uniform case.

Neither approach aims to reproduce the true astrophysical GRB population, which remains unmodeled. A strictly realistic population model should account for the cosmological correlations between distances, energy and observed durations. It is not the primary objective of this paper. Rather, the two together allow us to assess, first, whether any region of the plausible parameter space could produce a detectable memory signal, and second, how likely such configurations are to occur in a more realistic population. To avoid confusion between the names of the step-function models and the population models, we refer to the first population model as \textit{unweighted} and the second as \textit{realistic} (within the assumptions discussed above).

\begin{table}
\caption{Optimistic GRB population parameters and priors (for the realistic population)}
\label{tab: optimistic_population}
\squeezetable
\begin{ruledtabular}
\begin{tabular}{l c c c}
Parameter & Distribution & Prior & Range \\
\colrule
$\log_{10}(E_{\mathrm{\gamma, iso}}) [\mathrm{erg}]$ & Lognormal & $\mathcal{N}(52, 1)$ & $[46, 55]$ \\
$\log_{10}(T_{90}) [\mathrm{s}]$ (short) & Lognormal & $\mathcal{N}(0.0208, 0.606)$ & $[0.1, 500]$ \\
$\log_{10}(T_{90}) [\mathrm{s}]$ (long) & Lognormal & $\mathcal{N}(1.476, 0.435)$ & $[0.1, 500]$ \\
$\log_{10}(E_{\mathrm{K}}) [\mathrm{erg}]$ & Lognormal & $\mathcal{N}(52, 1)$ & $[50, 56]$ \\
$t_{\mathrm{dec}} [\mathrm{s}]$ & Normal & $\mathcal{N}(100, 1)$ & $[0.01, 1000]$ \\
$t_{\mathrm{break}}[\mathrm{day}]$ & Normal & $\mathcal{N}(1, 1)$ & $[0.1, 10]$ \\
$\log_{10}(r) [\mathrm{pc}]$ & Lognormal & $\mathcal{N}(9, 1)$ & $[7.60, 10]$ \\
$\theta~[\mathrm{rad}]$ & Uniform  & $[0, \pi/2]$ & $[0, \pi/2]$ \\
$\Gamma$ & Uniform  & $[100, 1000]$ & $[100, 1000]$ \\
\end{tabular}
\end{ruledtabular}
\end{table}

\subsection{Results}
\label{sec:results_population}
In this part, the SNR results for the two GRBs populations are presented, first for ET and then for LISA with the same methods explained in Sec.~\ref{sec:snr_and_case_study}.
\subsubsection{Einstein Telescope}
Based on the two generated populations of 10\,000 samples, we evaluate the SNR in ET for both the uniform and exponential models, and identify the GRB configurations that yield the best detectability.

\begin{figure}
    \centering
    \includegraphics[width=1\linewidth]{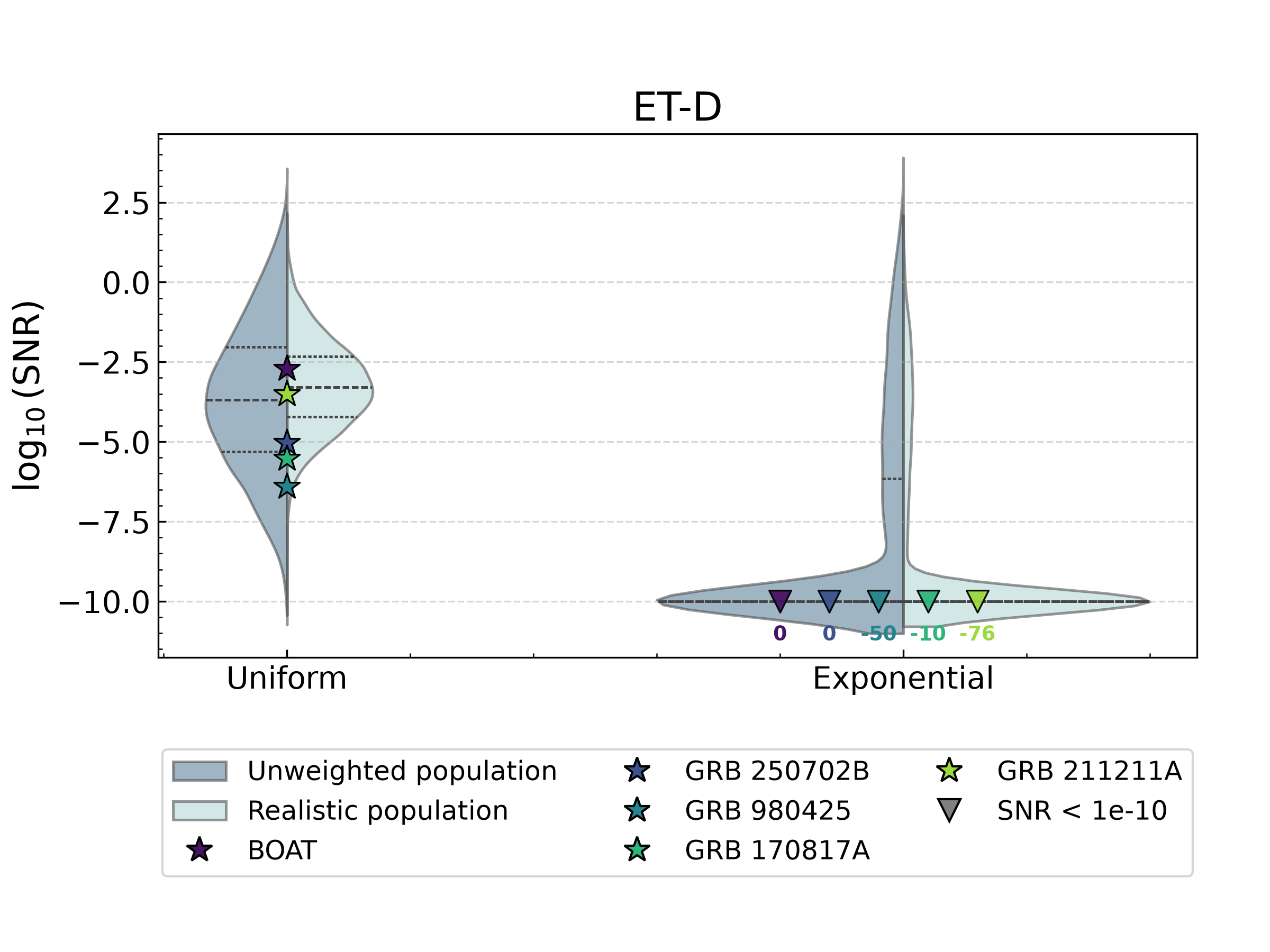}
    \caption{SNR histograms in ET for the uniform and exponential models, and for the unweighted and realistic GRB populations.}
    \label{fig:snr_histogram_ET}
\end{figure}

We first note in Fig.~\ref{fig:snr_histogram_ET} that both populations yield an overall low SNR in ET. For the unweighted population, the median SNR is $2.031 \times 10^{-4}$ for the uniform model and $4.078 \times 10^{-19}$ for
the exponential model. For the realistic population, the median SNR is $5.060 \times 10^{-4}$ for the uniform model and $8.661 \times 10^{-36}$ for the exponential model. As in the GW170817 case, the exponential model is more pessimistic than the uniform one, since its steeper high-frequency decay removes power from the region of ET's sensitive band. This effect is amplified in the realistic population, where the most extreme (and hence most detectable) joint configurations are strongly disfavored by the priors, so that the median SNR of the exponential model becomes negligible. Conversely, the realistic population's uniform-model median SNR is similar to that of the unweighted population, reflecting the fact that the priors affect the uniform model's SNR less strongly than the exponential one.

Considering each GRB individually, the majority of configurations yield a higher SNR under the uniform model, as expected. However, a subset of GRBs shows the opposite behavior, corresponding to bursts with a very short prompt duration, for which the exponential decay does not have time to remove significant power before ET's maximum sensitive frequency is reached. For the unweighted population, this subset is characterized by $T_{90} = 1.214^{+0.174}_{-0.150} \times 10^{-1}$~s, and for the realistic population by $T_{90} = 1.235^{+0.159}_{-0.124} \times 10^{-1}$~s.

The fraction of configurations whose SNR exceeds the ET detection threshold of 8 further illustrates the effect of the priors. For the unweighted population, 246 out of 10\,000 configurations are detectable under the uniform model, against 122 under the exponential model. For the realistic population, it drops to 17 out of 10\,000 under the uniform model, and only 6 under the exponential model.

\begin{figure*}
    \centering
    \includegraphics[width=1\linewidth]{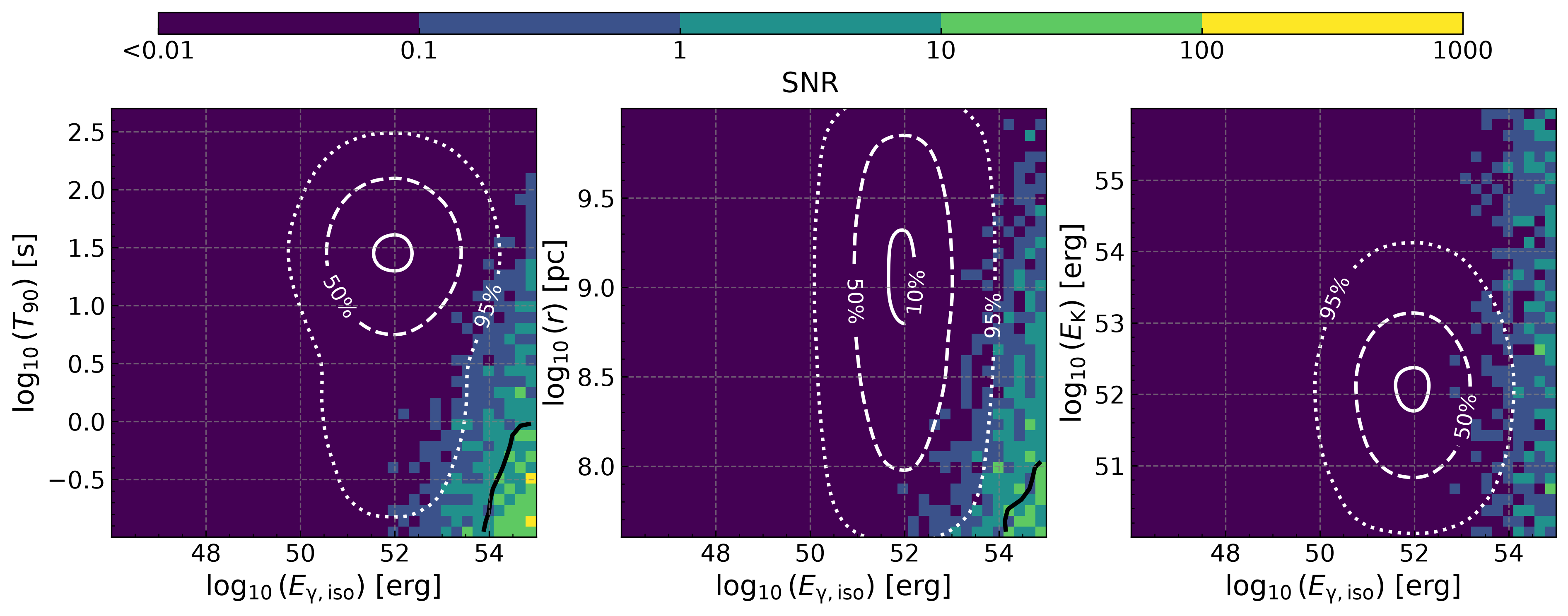}
    \caption{SNR heatmaps for ET obtained from the uniform model. The white contours represent the realistic population following the distributions in Table~\ref{tab: optimistic_population}, with the enclosed regions indicating the proportion of GRBs included within each curve (dotted: 95\%, dashed: 50\%, solid: 10\%). The black contour marks the SNR = 8 threshold.}
    \label{fig:heatmap_snr_1x3_E_GRB_uniform}
\end{figure*}

To visually illustrate how restricted the detectable parameter space is, we present SNR heatmaps in Fig.~\ref{fig:heatmap_snr_1x3_E_GRB_uniform}. These heatmaps demonstrate that high-SNR events are strictly confined to a narrow corner of the parameter space. Hence, only a negligible fraction of the population is loud enough to be observed by ET.

\begin{figure}
    \centering
    \includegraphics[width=1\linewidth]{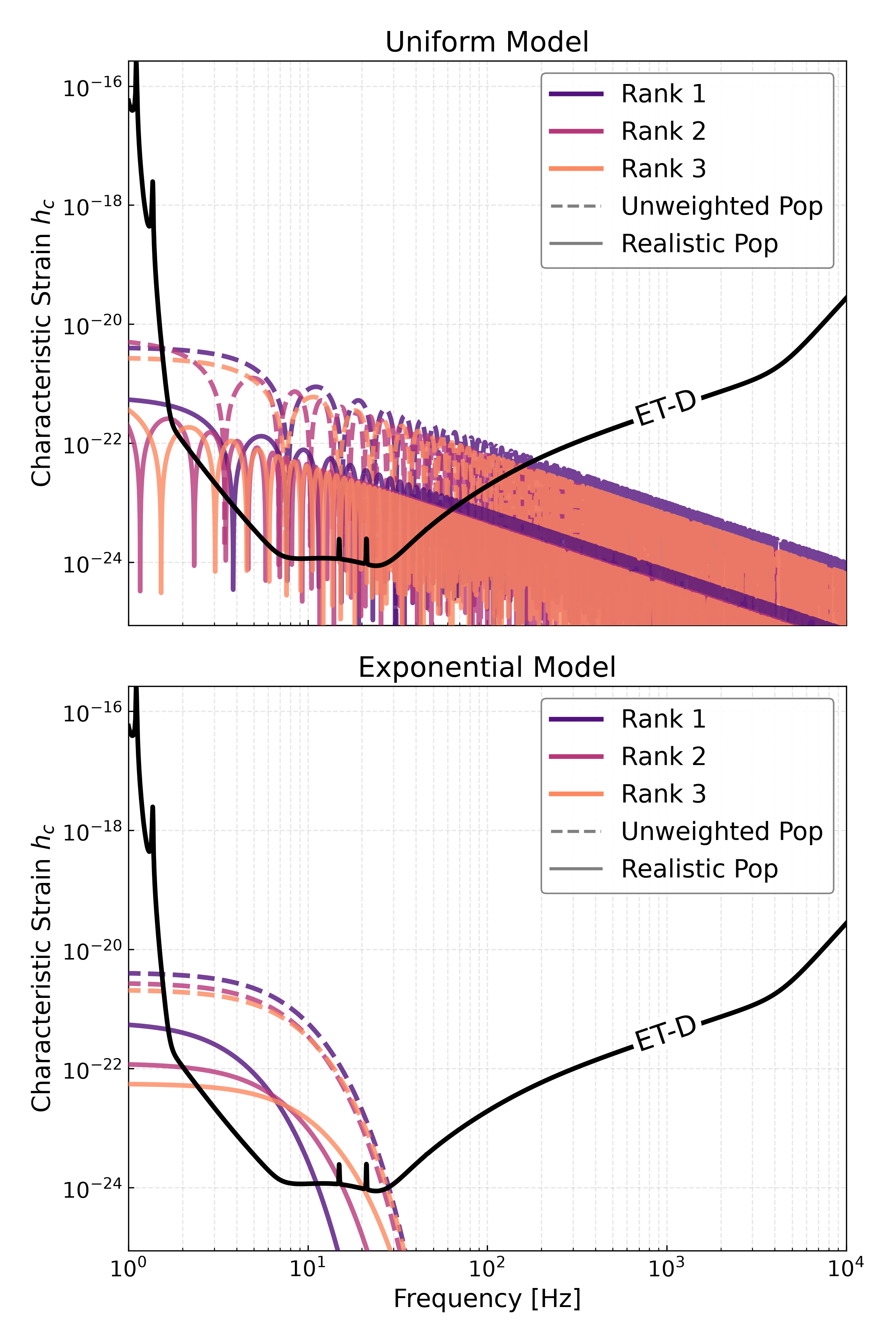}
    \caption{Three best candidates for a GRB memory detection with ET, for the unweighted and realistic populations. The characteristic strain of each GRB is shown
    for the uniform (\textit{top}) and exponential (\textit{bottom}) model, together with ET sensitivity curve (black) and the two population. Parameter values for each configuration are listed in Table~\ref{tab:model_stats_ET}.}
    \label{fig:top_candidates_hc_ET}
\end{figure}

The top three candidates for each model and population are shown in Fig.~\ref{fig:top_candidates_hc_ET}, and the corresponding parameter distributions for all detectable configurations are summarized in Table~\ref{tab:model_stats_ET}.

\begin{table*}
    \caption{Properties of the detectable GRBs ($\rho > 8$) in ET, for the unweighted and realistic populations and for the uniform and exponential rising-phase models.}
    \label{tab:model_stats_ET}
    \footnotesize
    \squeezetable
    \begin{ruledtabular}
    \begin{tabular}{lcccc}
        Parameter & \multicolumn{2}{c}{Unweighted population} & \multicolumn{2}{c}{Realistic population} \\
                  & Uniform Model & Exponential Model & Uniform Model & Exponential Model \\
        \colrule
        $E_{\gamma,\rm{iso}}$ [erg] & $(3.31^{+4.04}_{-2.43}) \times 10^{54}$ & $(2.79^{+3.83}_{-2.20}) \times 10^{54}$ & $(9.82^{+40.47}_{-4.24}) \times 10^{53}$ & $(8.84^{+12.29}_{-2.51}) \times 10^{53}$ \\
        $T_{90}$ [s]                & $0.30^{+0.79}_{-0.16}$                  & $0.16^{+0.09}_{-0.04}$                  & $0.34^{+0.63}_{-0.16}$                 & $0.18^{+0.06}_{-0.04}$                 \\
        $E_{\rm{K}}$ [erg]          & $(8.31^{+866.0}_{-8.19}) \times 10^{52}$ & $(7.45^{+932.1}_{-7.34}) \times 10^{52}$ & $(1.24^{+9.60}_{-1.10}) \times 10^{52}$ & $(5.56^{+25.59}_{-5.41}) \times 10^{52}$ \\
        $T_{\rm{aft}}$ [s]          & $(7.74^{+31.90}_{-5.74}) \times 10^{4}$ & $(7.14^{+31.05}_{-5.20}) \times 10^{4}$ & $(1.17^{+1.67}_{-0.93}) \times 10^{5}$ & $(2.20^{+1.65}_{-1.65}) \times 10^{5}$ \\
        $r$ [Mpc]                  & $119.6^{+288.4}_{-64.1}$                & $136.2^{+419.7}_{-80.0}$                & $83.3^{+87.9}_{-23.4}$                 & $114.4^{+110.7}_{-44.6}$               \\
        $\theta$ [rad]              & $0.74^{+0.51}_{-0.48}$                  & $0.70^{+0.57}_{-0.44}$                  & $0.63^{+0.43}_{-0.42}$                 & $0.41^{+0.25}_{-0.23}$                 \\
        $\Gamma$                    & $97.0^{+378.2}_{-75.0}$                 & $79.6^{+348.6}_{-59.7}$                 & $116.7^{+341.6}_{-98.8}$               & $66.4^{+142.4}_{-48.4}$                \\
    \end{tabular}
    \end{ruledtabular}
\end{table*}

For both populations and both models, the detectable configurations remain extreme in that they combine:
\begin{enumerate}

\item A prompt-emission energy comparable to, or approaching, that of the BOAT.
\item A very short duration (sub-second), characteristic of short GRBs typically associated with BNS mergers.
\item A luminosity distance of $\sim 100$~Mpc. 
\end{enumerate}
To date, no observed GRB combines all three of these properties simultaneously. This is also confirmed by the low SNR of the GRBs used to construct the population (see Fig.~\ref{fig:snr_histogram_ET}), as they lack the necessary extreme features.

Comparing the two populations, the realistic case systematically requires a less extreme combination of parameters to reach the detection threshold: the median energy of detectable bursts is lower by roughly a factor of three, compensated by a smaller required distance ($r\sim 120$~Mpc versus $\sim 83$~Mpc), consistent with the priors of Table~\ref{tab: optimistic_population} disfavouring the most extreme energies. Overall, the realistic population produces both a lower typical SNR and a much smaller fraction of detectable configurations than the unweighted population, and the gap between the uniform and exponential models widens once physically motivated priors are applied.

\subsubsection{LISA}
\begin{figure}
    \centering
    \includegraphics[width=1\linewidth]{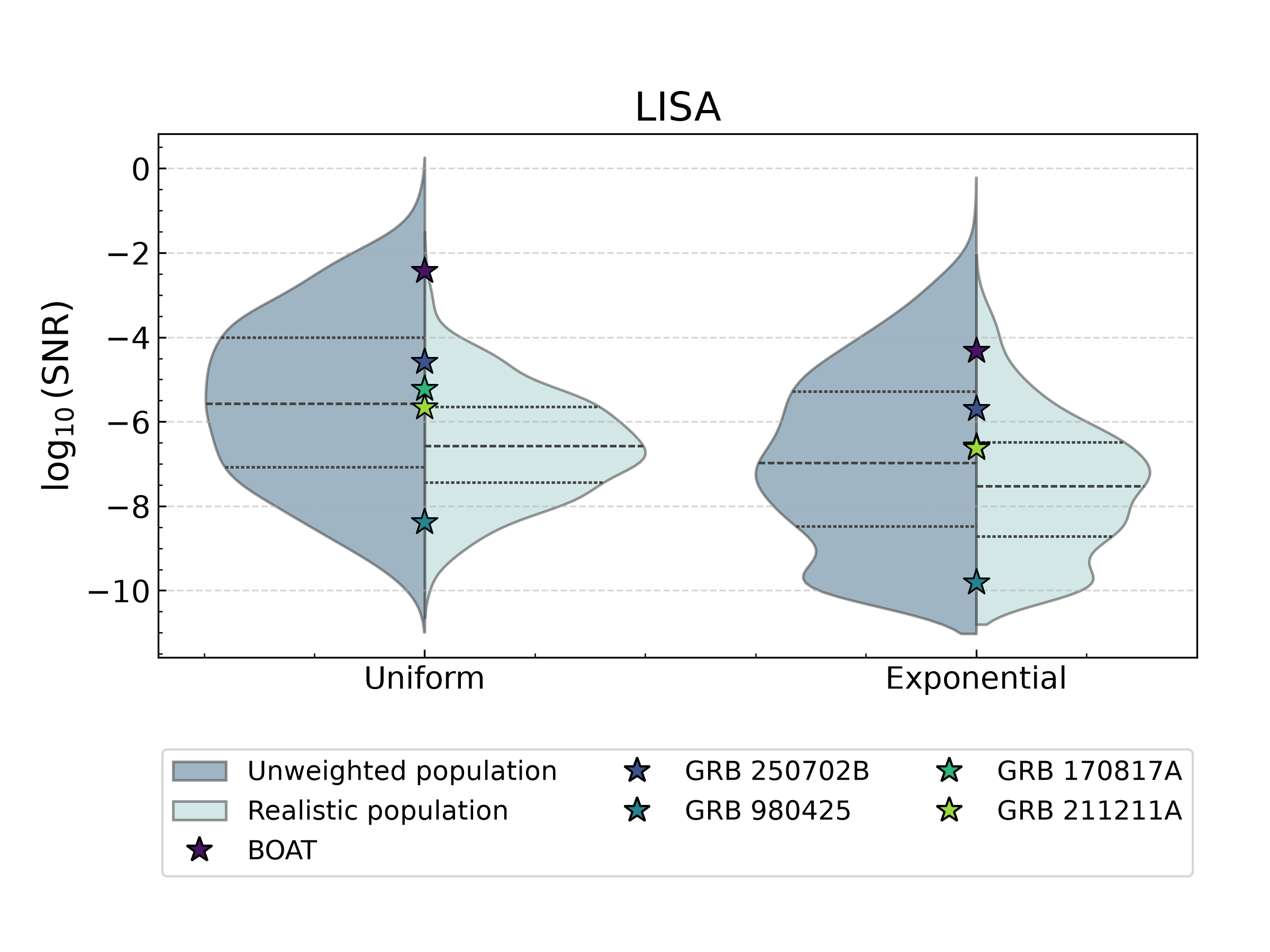}
    \caption{SNR histograms in LISA for the uniform and exponential models, and for the unweighted  and realistic GRB populations.}
    \label{fig:results_snr_grb_LISA}
\end{figure}

After projecting the population onto the LISA response, no GRB configuration yields a sufficiently high SNR to conclude that detection is feasible. As shown in Fig.~\ref{fig:results_snr_grb_LISA}, the median SNR is very low for all models and both populations. For the unweighted population, the median SNR is $2.701^{+3.579\times10^{2}}_{-2.679}\times10^{-6}$ for the uniform model and $1.071^{+203.5}_{-1.066}\times10^{-7}$ for the exponential model. For the realistic population, the median SNR is
$2.679^{+51.65}_{-2.531}\times10^{-7}$ for the uniform model and
$3.009^{+98.79}_{-2.963}\times10^{-8}$ for the exponential model. Moreover, the specific GRBs defining our population bounds do not possess a high enough SNR to be detectable in LISA. The GRB with the highest SNR is listed in Table~\ref{tab:top10_candidates_LISA} and does not exceed $\rho = 0.2$.

\begin{figure*}
    \centering
    \includegraphics[width=1\linewidth]{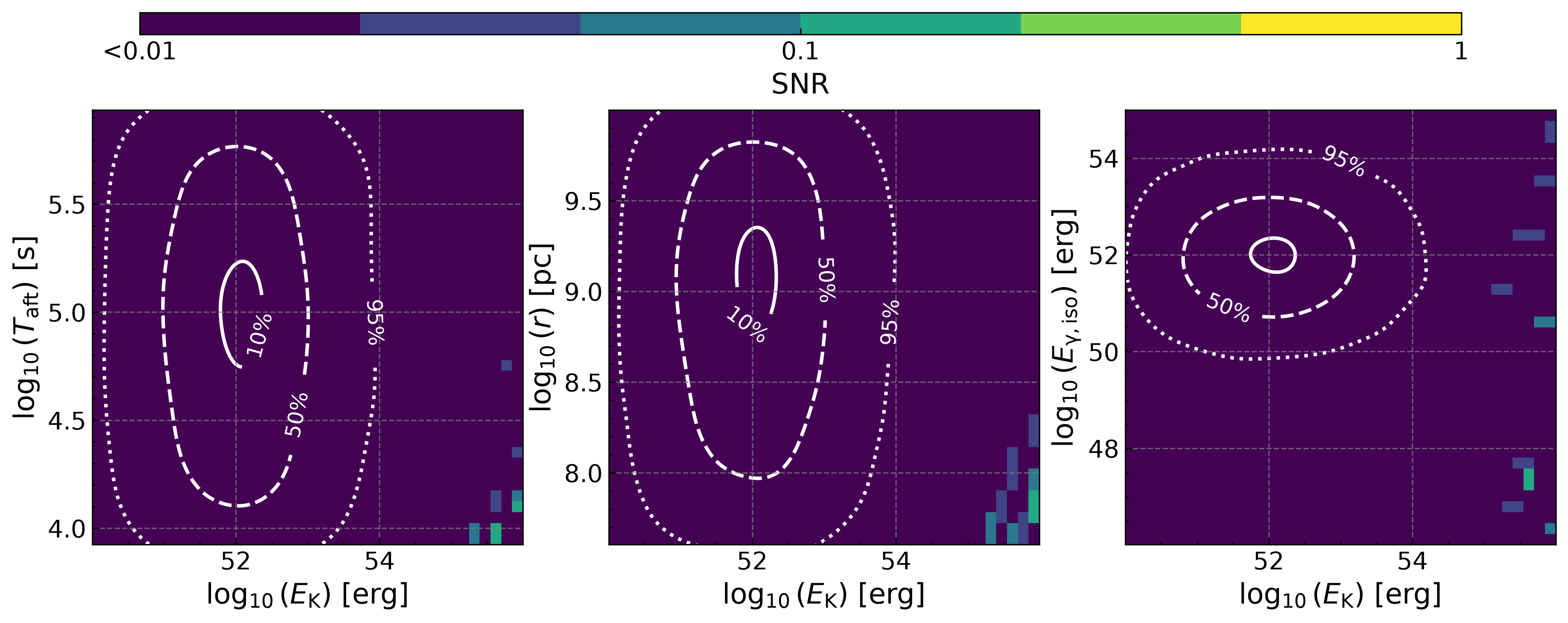}
    \caption{SNR heatmaps for LISA obtained from the uniform model. The white contours represent the realistic population following the distributions in Table~\ref{tab: optimistic_population}, with the enclosed regions indicating the proportion of GRBs included within each curve (dotted: 95\%, dashed: 50\%, solid: 10\%).}
    \label{fig: heatmap_snr_1x3_E_aft_uniform}
\end{figure*}

The widespread lack of detectability is confirmed by the LISA SNR heatmaps in Fig.~\ref{fig: heatmap_snr_1x3_E_aft_uniform}. Unlike ET, the heatmaps show that the parameter space remains constrained to sub-threshold events.

Two main factors explain these low SNR values:
\begin{enumerate}
    \item The amplitude of the GRB memory effect is too weak for such parameter configurations.
    \item The characteristic timescale of the afterglow phase pushes its counterpart outside the optimal frequency band of LISA.
\end{enumerate}

\ignore{Two factors account for these low SNR values. First, applying the full LISA response more accurately captures the detector behavior but reduces the effective amplitude (roughly of two order of magnitude) of the signal compared to the idealized response estimate used for ET. Second, the duration of the afterglow emission may not be well matched to the frequency band in which LISA is most sensitive, so that the bulk of the signal power falls outside the instrument's optimal sensitivity range.}

To isolate the impact of the timescale mismatch, we tested an extreme GRB configuration (extreme in the sense that each individual parameter lies at the edge of what has already been observed in the realistic population) with a very bright prompt and afterglow emission: a prompt isotropic energy of $10^{55}\,\mathrm{erg}$ with a duration $T_{90}$ of $100\,\mathrm{s}$, an afterglow isotropic energy of $10^{56}\,\mathrm{erg}$, and a bulk Lorentz factor of $1000$, placed at a distance of $40\,\mathrm{Mpc}$ and observed at a viewing angle of $0.001\,\mathrm{rad}$ from the jet axis. We found that the SNR is maximized for an afterglow duration of approximately $100\,\mathrm{s}$, reaching a value just below $\rho = 10$. Nevertheless, more realistic afterglows, which typically last longer than one day, fall outside this favorable frequency range and therefore remain undetectable by LISA.

\section{Discussion \& Conclusion}
\label{sec:discussion}

In this paper, we have quantified both the nonlinear and linear GW memory contributions generated during a BNS merger and their SNR in ET and LISA. For the first time, we combined all expected sources of memory effects and used toy-models and simulations to a system similar to GW170817, spanning the $10^{-5}$~Hz to $10^{4}$~Hz frequency band probed by these two next-generation detectors. We found that the combined memory signal would be detectable by ET with an SNR of $10.7$. Its detectability is dominated by the nonlinear memory component. Among the linear contributions, the dynamical ejecta provides the highest strain, but its impact remains minimal. The memory induced by the disk wind ejecta and the prompt emission from GRB170817A are negligible. Extended emissions including the KN and the afterglow part of the GRB evolve over timescales that are too long to be effectively detected by ET. For the neutrino counterpart, its detectability is sensitive to the fate of the remnant. In the optimistic scenario of a long-lived stable remnant, the extended neutrino flux is larger than the dynamical ejecta ($\rho \sim 0.6$), whereas a prompt collapse renders the signal weaker ($\rho \sim 0.02$). Similarly, we can expect that the nature of the remnant will lower the mass of unbound matter expelled through dynamical and disk wind ejecta~\cite{Metzger:2019zeh, Kawaguchi_2020}, directly affecting their memory amplitude. For LISA, our results predict no detectable signal.

Our results are consistent with the recent GRMHD study of Bamber et al. (2026)~\cite{Bamber_2026} who investigated the total memory signal from BNS over the $\sim 25~\rm{ms}$ surrounding the merger. They found that the nonlinear (named GW null in their work) dominates the total signal for all and the most extreme magnetic field configurations, with linear contributions from electromagnetic radiation, neutrinos, and ejecta reaching at most $\sim 15 \%$ of the total memory in typical cases. This hierarchy matches our finding, with the linear contributions remaining subdominant compared with the nonlinear.

Recent analysis of GWTC-4 estimates the local volumetric rate of GW170817-like systems at $53^{+176}_{-49}\,\mathrm{Gpc}^{-3}\,\mathrm{yr}^{-1}$, representing barely half of the total BNS merger rate~\cite{Fishbach:2026}. Since ET is expected to detect $\sim 10^{4}$ BNS mergers per year~\cite{Maggiore_2020}, this translates to thousands of GW170817-like events annually. If the linear memory of a single event remains too faint, statistically stacking these signals is an approach to attempt to detect their signal~\cite{Lasky_2016, Lopez:2023aja}.

Expanding our analysis beyond a single event, we have also investigated the memory produced by GRBs across two synthetic populations: unweighted and realistic. For ET, we found that detecting the linear memory from the prompt phase is theoretically possible, but needs highly favorable source properties. Specifically, it requires nearby GRB events with sub-second burst durations and isotropic prompt energies at least higher than the brightest observed GRBs. For LISA, our population study demonstrated that no GRB configurations at realistic distances yielded a sufficiently high SNR to claim a detection. Moreover, applying observationally motivated priors (realistic population) reduces the number of detectable events compared to an unweighted exploration of the parameter space. 

The detectability results highlight a strong sensitivity to the chosen temporal profile in our toy models. The contrast in SNR between the uniform and exponential rising-phase models stems directly from their behavior in frequency domain. The exponential model decays much more steeply at frequencies above the inverse of the characteristic ejection timescale. By comparison, the uniform model yields a shallower characteristic strain decay and provides a more optimistic SNR for slow processes whose knee frequencies fall below the detector optimal band. Nevertheless, the SNR estimations of both models align for very rapid ejection mechanisms, such as the dynamical ejecta ($\tau \sim 1$~ms). In such cases, the detector only probes the low-frequency plateau of the characteristic strain.

Furthermore, isolating the distinct signatures that comprise the linear memory presents another challenge following the detection. In the time domain, different linear memory effects can be easily distinguished provided their characteristic timescales do not significantly overlap. However, in the frequency domain, it becomes complex to pinpoint which specific astrophysical counterpart produced the observed memory effect, as the low-frequency contributions of multiple components blend together. By establishing comprehensive physical priors on the expected timescales and relative amplitudes of these diverse signatures, the global framework presented in this paper can serve as a first tool to help disentangle and identify these different phenomena in future observations.

From the modeling of GRB memory, the model neglects alternative energy injection mechanisms. In afterglow lightcurves, a bump is sometimes present and is thought to be related to an additional energy injection from the central engine. As discussed in \citet{EInjection_GRBmem}, the energy injection would produce a memory effect. However, because such phenomena remain actively debated and poorly constrained, we excluded this component from our analysis. 

Improvements can be made to the toy-models by moving away from the point-particle approximation and instead modeling realistic density and velocity fields for the ejecta and GRB jets. Some results have already been obtained for uniform or structured jet geometries~\cite{Birnholtz_2013, Brabant:2026}. For an off-axis observer, the memory amplitude decreases as the jet opening angle increases because the energy inside the jet is spread over a larger solid angle, reducing the energy density near the observer's line of sight~\cite{Brabant:2026}. For an on-axis observer, the point-particle model predicts zero memory due to anti-beaming, but a finite-width jet produces a non-zero signal because off-axis regions of the jet contribute emission that is no longer perfectly canceled~\cite{Birnholtz_2013, Brabant:2026}.\\

Finally measuring the memory effect will be challenged by parameter degeneracies and the dominance of the nonlinear component. Overcoming the former is possible thanks to joint multimessenger observations. Electromagnetic counterparts  are indeed necessary to independently constrain the system parameters. A method including only the nonlinear memory has been used to break the distance-inclination degeneracy~\cite{dist_incl_breaking}, and a similar approach involving the linear memory in a BNS study is a promising improvement. 

In the absence of an electromagnetic counterpart, the memory signal offers a compelling science case. As next-generation detectors come online, a challenge will be to accurately model and subtract the dominant nonlinear memory from the data. Analyzing the residuals to isolate the linear memory will open a novel avenue. The residual memory signal provides evidence of energy dissipation beyond gravitational radiation. Ultimately, successfully disentangling the linear and nonlinear memory contributions could yield insights into the radiation and mass-ejection processes of CBCs through a new channel.  

\begin{acknowledgments}
HI thanks the Belgian Federal Science Policy Office (BELSPO) for the provision of financial support in the framework of the PRODEX Programme of the European Space Agency (ESA) under contract number PEA4000144253. The authors thank Diego Blas for discussions that motivated this project and for the initial idea that inspired this work. MP and TB thank J.R Cudell for the useful discussions. MB acknowledges the Department of Physics and Earth Science of the University of Ferrara for the financial support through the FIRD 2025 grant.
\end{acknowledgments}
\appendix
\section{Best GRB configurations in ET and LISA}
Below are Tables~\ref{tab:top10_candidates_ET} and~\ref{tab:top10_candidates_LISA}, which contain the top five configurations yielding the highest SNR for ET and LISA, respectively, for both the unweighted and realistic populations, and for the uniform and exponential step function models.
\begin{table*}
    \caption{Top configurations yielding the highest SNR in ET, for the unweighted and realistic populations, and for the uniform and exponential rising-phase models.}
    \label{tab:top10_candidates_ET}
    \footnotesize
    \squeezetable
    \begin{ruledtabular}
    \begin{tabular}{ccccccccc}
        {Rank} & {$\rho$} & {$E_{\gamma,\rm{iso}}$ [$10^{54}$ erg]} & {$T_{90}$ [s]} & {$E_{\rm{K}}$ [$10^{52}$ erg]} & {$T_{\rm{aft}}$ [days]} & {$r$ [Mpc]} & {$\theta$ [rad]} & {$\Gamma$} \\
        \colrule
        \multicolumn{9}{c}{{Uniform Model -- Unweighted Population}} \\
        \colrule
        1  & 659.78 & 7.55 & 0.13 & 0.02    & 3.16  & 61.53  & 0.23 & 54.66  \\
        2  & 447.06 & 7.42 & 0.29 & 0.21    & 3.71  & 43.11  & 0.19 & 58.69  \\
        3  & 428.89 & 4.95 & 0.13 & 3.62    & 7.34  & 50.16  & 0.87 & 647.55 \\
        4  & 366.98 & 9.07 & 0.20 & 76.57   & 1.77  & 56.53  & 1.45 & 191.40 \\
        5  & 355.61 & 6.42 & 0.12 & 1064.09 & 0.68  & 97.99  & 0.47 & 80.60  \\
        \ignore{6  & 332.97 & 6.37 & 0.27 & 123.65  & 0.47  & 54.32  & 0.25 & 240.88 \\
        7  & 328.53 & 6.03 & 0.23 & 0.65    & 3.42  & 64.10  & 0.10 & 384.43 \\
        8  & 304.48 & 9.70 & 0.34 & 1406.44 & 0.49  & 67.91  & 0.35 & 15.51  \\
        9  & 298.54 & 8.47 & 0.14 & 0.29    & 0.25  & 104.16 & 1.08 & 16.20  \\
        10 & 285.82 & 5.83 & 0.11 & 0.18    & 9.27  & 118.80 & 0.66 & 12.87  \\}
        \colrule
        \multicolumn{9}{c}{{Uniform Model -- Realistic Population}} \\
        \colrule
        1  & 52.78  & 1.51 & 0.26 & 0.03    & 2.77  & 83.28  & 0.23 & 53.23  \\
        2  & 31.70  & 6.46 & 0.86 & 1.65    & 0.15  & 166.81 & 0.63 & 561.66 \\
        3  & 29.71  & 3.28 & 0.66 & 2.52    & 1.64  & 73.88  & 1.45 & 189.05 \\
        4  & 27.43  & 5.67 & 1.60 & 12.04   & 0.54  & 94.10  & 0.35 & 15.52  \\
        5  & 24.57  & 2.25 & 0.34 & 0.22    & 0.28  & 154.84 & 1.08 & 16.20  \\
        \ignore{6  & 22.16  & 7.18 & 0.40 & 0.08    & 1.35  & 392.43 & 1.33 & 124.23 \\
        7  & 21.88  & 0.63 & 0.29 & 0.71    & 6.73  & 62.09  & 0.88 & 633.17 \\
        8  & 21.45  & 0.98 & 0.24 & 9.88    & 0.73  & 145.52 & 0.47 & 79.60  \\
        9  & 17.55  & 0.79 & 0.17 & 0.18    & 9.06  & 176.78 & 0.66 & 12.85  \\
        10 & 17.50  & 1.35 & 1.25 & 0.19    & 3.27  & 47.03  & 0.18 & 57.25  \\}
        \colrule
        \multicolumn{9}{c}{{Exponential Model -- Unweighted Population}} \\
        \colrule
        1  & 770.11 & 7.55 & 0.13 & 0.02    & 3.16  & 61.53  & 0.23 & 54.66  \\
        2  & 484.04 & 4.95 & 0.13 & 3.62    & 7.34  & 50.16  & 0.87 & 647.55 \\
        3  & 426.84 & 6.42 & 0.12 & 1064.09 & 0.68  & 97.99  & 0.47 & 80.60  \\
        4  & 354.95 & 5.83 & 0.11 & 0.18    & 9.27  & 118.80 & 0.66 & 12.87  \\
        5  & 347.36 & 2.77 & 0.11 & 13.66   & 2.62  & 56.08  & 0.67 & 117.79 \\
        \ignore{6  & 314.01 & 8.47 & 0.14 & 0.29    & 0.25  & 104.16 & 1.08 & 16.20  \\
        7  & 225.88 & 9.07 & 0.20 & 76.57   & 1.77  & 56.53  & 1.45 & 191.40 \\
        8  & 212.03 & 3.95 & 0.16 & 22.36   & 0.99  & 58.41  & 1.04 & 170.99 \\
        9  & 174.69 & 5.48 & 0.10 & 5418.14 & 3.74  & 294.34 & 0.36 & 586.13 \\
        10 & 168.46 & 3.85 & 0.11 & 6.55    & 3.40  & 162.02 & 0.70 & 26.05  \\}
        \colrule
        \multicolumn{9}{c}{{Exponential Model -- Realistic Population}} \\
        \colrule
        1  & 19.74  & 1.51 & 0.26 & 0.03    & 2.77  & 83.28  & 0.23 & 53.23  \\
        2  & 16.38  & 4.53 & 0.15 & 25.82   & 0.26  & 57.02  & 0.01 & 19.31  \\
        3  & 15.11  & 0.71 & 0.10 & 52.46   & 3.30  & 418.18 & 0.35 & 577.41 \\
        4  & 14.28  & 0.79 & 0.17 & 0.18    & 9.06  & 176.78 & 0.66 & 12.85  \\
        5  & 11.30  & 0.33 & 0.19 & 1.24    & 2.31  & 72.94  & 0.67 & 116.72 \\
        \ignore{6  & 9.70   & 0.98 & 0.24 & 9.88    & 0.73  & 145.52 & 0.47 & 79.60  \\
        7  & 6.43   & 0.63 & 0.29 & 0.71    & 6.73  & 62.09  & 0.88 & 633.17 \\
        8  & 6.34   & 0.27 & 0.13 & 2.80    & 0.41  & 217.73 & 0.76 & 344.81 \\
        9  & 5.81   & 0.06 & 0.14 & 3.28    & 0.62  & 59.51  & 0.32 & 49.34  \\
        10 & 5.76  & 0.08 & 0.11 & 2.41    & 4.30  & 102.62 & 0.69 & 34.45  \\}
    \end{tabular}
    \end{ruledtabular}
\end{table*}

\begin{table*}
    \caption{Top configurations yielding the highest SNR in LISA, for the unweighted and realistic populations, and for the uniform and exponential rising-phase models.}
    \label{tab:top10_candidates_LISA}
    \footnotesize
    \squeezetable
    \begin{ruledtabular}
    \begin{tabular}{ccccccccc}
        {Rank} & {$\rho$} & {$E_{\gamma,\rm{iso}}$ [$10^{54}$ erg]} & {$T_{90}$ [s]} & {$E_{\rm{K}}$ [$10^{52}$ erg]} & {$T_{\rm{aft}}$ [days]} & {$r$ [Mpc]} & {$\theta$ [rad]} & {$\Gamma$} \\
        \colrule
        \multicolumn{9}{c}{{Uniform Model -- Unweighted Population}} \\
        \colrule
        1  & 0.19                   & $3.21 \times 10^{-4}$  & 113.24     & 9.68                   & 0.15       & 57.69      & 1.47  & 16.78      \\
        2  & 0.16                   & $1.49 \times 10^{-7}$  & 361.52     & 4.27                   & 0.12       & 52.72      & 1.15  & 463.69     \\
        3  & 0.09                   & $1.50 \times 10^{-3}$  & 1.92       & 2.47                   & 0.11       & 41.98      & 1.26  & 860.27     \\
        4  & 0.08                   & $1.83 \times 10^{-8}$  & 0.92       & 7.54                   & 0.16       & 95.32      & 1.48  & 278.50     \\
        5  & 0.04                   & 0.02                   & 259.67     & 3.57                   & 0.14       & 92.14      & 0.83  & 159.87     \\
        \ignore{6  & 0.04                   & 0.33                   & 231.70     & 9.34                   & 0.25       & 176.51     & 0.13  & 35.89      \\
        7  & 0.04                   & $5.86 \times 10^{-7}$  & 6.59       & 2.93                   & 0.17       & 76.47      & 0.94  & 30.58      \\
        8  & 0.03                   & 0.03                   & 11.29      & 5.98                   & 0.68       & 51.15      & 0.98  & 47.94      \\
        9  & 0.03                   & 2.43                   & 4.93       & 9.88                   & 1.03       & 54.94      & 0.88  & 115.95     \\
        10 & 0.02                   & $5.21 \times 10^{-8}$  & 81.10      & 2.17                   & 0.12       & 141.33     & 1.30  & 44.99      \\}
        \colrule
        \multicolumn{9}{c}{{Uniform Model -- Realistic Population}} \\
        \colrule
        1  & $7.16 \times 10^{-3}$  & $9.42 \times 10^{-3}$  & 62.52      & 0.53                   & 0.16       & 75.78      & 1.47  & 16.78      \\
        2  & $2.29 \times 10^{-3}$  & 0.30                   & 17.55      & 1.06                   & 1.04       & 70.73      & 0.89  & 114.58     \\
        3  & $1.96 \times 10^{-3}$  & 0.05                   & 58.06      & $3.42 \times 10^{-4}$  & 0.14       & 60.31      & 0.04  & 82.98      \\
        4  & $1.02 \times 10^{-3}$  & $7.19 \times 10^{-4}$  & 153.96     & 0.04                   & 0.13       & 66.31      & 1.15  & 462.59     \\
        5  & $9.43 \times 10^{-4}$  & 0.09                   & 97.80      & 0.35                   & 0.29       & 264.87     & 0.13  & 34.78      \\
        \ignore{6  & $7.70 \times 10^{-4}$  & 0.04                   & 221.22     & $7.08 \times 10^{-3}$  & 0.31       & 207.48     & 0.05  & 267.42     \\
        7  & $7.00 \times 10^{-4}$  & $3.47 \times 10^{-4}$  & 2.40       & 0.92                   & 0.25       & 1157.99    & 0.96  & 38.53      \\
        8  & $6.97 \times 10^{-4}$  & $1.25 \times 10^{-4}$  & 6.39       & 0.11                   & 0.18       & 141.75     & 1.48  & 275.51     \\
        9  & $6.39 \times 10^{-4}$  & 0.01                   & 10.90      & 0.02                   & 0.12       & 44.51      & 1.26  & 858.33     \\
        10 & $3.61 \times 10^{-4}$  & 0.22                   & 1.11       & $1.47 \times 10^{-4}$  & 0.10       & 352.69     & 0.02  & 298.88     \\}
        \colrule
        \multicolumn{9}{c}{{Exponential Model -- Unweighted Population}} \\
        \colrule
        1  & 0.06                   & $1.49 \times 10^{-7}$  & 361.52     & 4.27                   & 0.12       & 52.72      & 1.15  & 463.69     \\
        2  & 0.01                   & 1.91                   & 115.63     & 0.05                   & 0.37       & 82.82      & 0.27  & 214.81     \\
        3  & $7.75 \times 10^{-3}$  & 8.99                   & 174.46     & $8.79 \times 10^{-3}$  & 0.30       & 80.02      & 1.46  & 537.78     \\
        4  & $6.13 \times 10^{-3}$  & $1.84 \times 10^{-6}$  & 12.32      & 6.85                   & 0.48       & 171.51     & 0.30  & 348.36     \\
        5  & $5.12 \times 10^{-3}$  & 0.08                   & 2.74       & 2.70                   & 0.17       & 256.05     & 0.34  & 222.17     \\
        \ignore{6  & $4.50 \times 10^{-3}$  & 0.03                   & 11.29      & 5.98                   & 0.68       & 51.15      & 0.98  & 47.94      \\
        7  & $3.93 \times 10^{-3}$  & 1.67                   & 0.27       & $1.29 \times 10^{-4}$  & 0.10       & 243.33     & 0.02  & 302.59     \\
        8  & $3.85 \times 10^{-3}$  & $3.21 \times 10^{-4}$  & 113.24     & 9.68                   & 0.15       & 57.69      & 1.47  & 16.78      \\
        9  & $2.45 \times 10^{-3}$  & $4.47 \times 10^{-8}$  & 0.45       & 9.85                   & 0.22       & 1012.89    & 0.95  & 39.83      \\
        10 & $2.39 \times 10^{-3}$  & 1.64                   & 37.00      & 0.01                   & 0.24       & 52.27      & 0.31  & 64.60      \\}
        \colrule
        \multicolumn{9}{c}{{Exponential Model -- Realistic Population}} \\
        \colrule
        1  & $1.44 \times 10^{-3}$  & 0.05                   & 58.06      & $3.42 \times 10^{-4}$  & 0.14       & 60.31      & 0.04  & 82.98      \\
        2  & $1.41 \times 10^{-3}$  & 0.71                   & 10.81      & $1.22 \times 10^{-3}$  & 0.56       & 182.05     & 0.06  & 31.84      \\
        3  & $8.33 \times 10^{-4}$  & 0.24                   & 63.21      & $2.02 \times 10^{-3}$  & 0.41       & 119.29     & 0.26  & 215.01     \\
        4  & $6.18 \times 10^{-4}$  & 1.01                   & 27.14      & $1.80 \times 10^{-4}$  & 0.15       & 248.00     & 0.20  & 46.70      \\
        5  & $5.83 \times 10^{-4}$  & 1.95                   & 327.44     & 0.02                   & 1.36       & 103.99     & 1.07  & 58.38      \\
        \ignore{6  & $4.28 \times 10^{-4}$  & 0.22                   & 1.11       & $1.47 \times 10^{-4}$  & 0.10       & 352.69     & 0.02  & 298.88     \\
        7  & $4.05 \times 10^{-4}$  & 0.13                   & 108.70     & $1.57 \times 10^{-3}$  & 0.46       & 92.60      & 0.40  & 21.98      \\
        8  & $3.60 \times 10^{-4}$  & $7.19 \times 10^{-4}$  & 153.96     & 0.04                   & 0.13       & 66.31      & 1.15  & 462.59     \\
        9  & $3.36 \times 10^{-4}$  & 5.00                   & 47.34      & $6.49 \times 10^{-5}$  & 1.87       & 1003.28    & 0.16  & 22.35      \\
        10 & $3.25 \times 10^{-4}$  & 0.43                   & 117.74     & $6.05 \times 10^{-3}$  & 2.45       & 47.49      & 0.27  & 39.68      \\}
    \end{tabular}
    \end{ruledtabular}
\end{table*}

\clearpage
\bibliography{ref}

\end{document}